\documentclass[preprint,12pt,sort&compress]{elsarticle}
\usepackage{graphicx}
\usepackage{amsbsy,amsmath,amsthm,latexsym,amssymb}
\usepackage{color,ulem}
\usepackage{lineno}
\newcommand{\bc}{\begin{center}}
\newcommand{\ec}{\end{center}}
\newcommand{\bfr}{\begin{flushright}}
\newcommand{\efr}{\end{flushright}}

\newcommand{\no}{\noindent}
\newcommand{\be}{\begin{enumerate}}
\newcommand{\ee}{\end{enumerate}}
\newcommand{\bi}{\begin{itemize}}
\newcommand{\ei}{\end{itemize}}
\newcommand{\bd}{\begin{description}}
\newcommand{\ed}{\end{description}}

\newcommand{\eeq}{\end{equation}}
\newcommand{\bea}{\begin{eqnarray}}
\newcommand{\eea}{\end{eqnarray}}

\newcommand{\bfi}{\begin{figure}}
\newcommand{\efi}{\end{figure}}
\newcommand{\bay}{\begin{array}{l}}
\newcommand{\eay}{\end{array}}

\newcommand{\del}{\delta}
\newcommand{\Del}{\Delta}

\newcommand{\Cel}{$^\circ$C~}  
\newcommand{\cref}[1]{(\ref{#1})}   

\newcommand{\ie}{\textit{i.e.}~}
\newcommand{\ca}{\textit{ca.}~}
\newcommand{\cf}{\textit{cf.}~}
\newcommand{\eg}{\textit{e.g.}~}
\newcommand{\via}{\textit{via}~}
\newcommand{\viz}{\textit{viz.}~}
\newcommand{\vs}{\textit{vs.}~}
\newcommand{\eqname}{Eq.~}
\newcommand{\eqsname}{Eqs.~}
\newcommand{\eqnames}{Eqs.~}
\newcommand{\figname}{Fig.~}
\newcommand{\figsname}{Figs.~}
\newcommand{\fignames}{Figs.~}
\newcommand{\tabname}{Table~}
\newcommand{\secname}{Section~}
\renewcommand{\thefootnote} 

\begin{document}
\baselineskip 26pt

\mbox{ } \vskip 22mm

\begin{center}

{\Large {\bf  C--S--H gel densification: the impact of the nanoscale on self desiccation and sorption isotherms}}

\vskip 7mm

Enrico Masoero, Gianluca Cusatis, Giovanni Di Luzio


%
%
%
\vskip 30mm


\vskip 25mm

\today

Corresponding Author

Enrico Masoero










\end{center}
\pagestyle{empty} \newpage

\pagestyle{plain}

\begin{center}
{\Large {\bf  C--S--H gel densification: the impact of the nanoscale on self desiccation and sorption isotherms}}
\\[9mm]
{\large {\bf Enrico Masoero$^{\mathrm{a,b}}$, \bf Gianluca Cusatis$^{\mathrm{c}}$, and Giovanni Di Luzio$^{\mathrm{b}}$
}}

\vskip 6mm

{\footnotesize $^{\mathrm{a}}$ School of Engineering, Newcastle University, NE1 7RU, Newcastle upon Tyne, U.K.}\\[1mm]

{\footnotesize $^{\mathrm{b}}$ Department of Civil and Environmental Engineering, Politecnico di Milano, Piazza Leonardo da Vinci 32,
20133 Milan, Italy}\\[1mm]

{\footnotesize $^{\mathrm{c}}$ Department of Civil and Environmental Engineering, Northwestern University, Evanston (IL), U.S.A.}\\[1mm]


\end{center}


\bigskip
\no {\bf   Abstract}\,
\vskip 3.5mm
{\small
The relationship between humidity and water content in a hydrating cement paste is largely controlled by the nanostructure of the C--S--H gel. Current hydration models do not describe this nanostructure, thus sorption isotherms and self-desiccation are given as constitutive inputs instead of being predicted from microstructural evolution. To address this limitation, this work combines a C--S--H gel description from nanoscale simulations with evolving capillary pore size distributions from a simple hydration model. Results show that a progressive densification of the C--S--H gel must be considered in order to explain the self-desiccation of low-alkali pastes. The impact of C--S--H densification on the evolution of microstructure and sorption isotherms is then discussed, including the effect of water-to-cement ratio, cement powder fineness, and curing temperature. Overall, this work identifies an area where nanoscale simulations can integrate larger-scale models of cement hydration and poromechanics.

}

%
%
\vskip 3mm \noindent

\noindent
\bigskip


\section{Introduction}\label{secIntro}

\begin{figure}
\centerline{\includegraphics[width=0.5\textwidth] {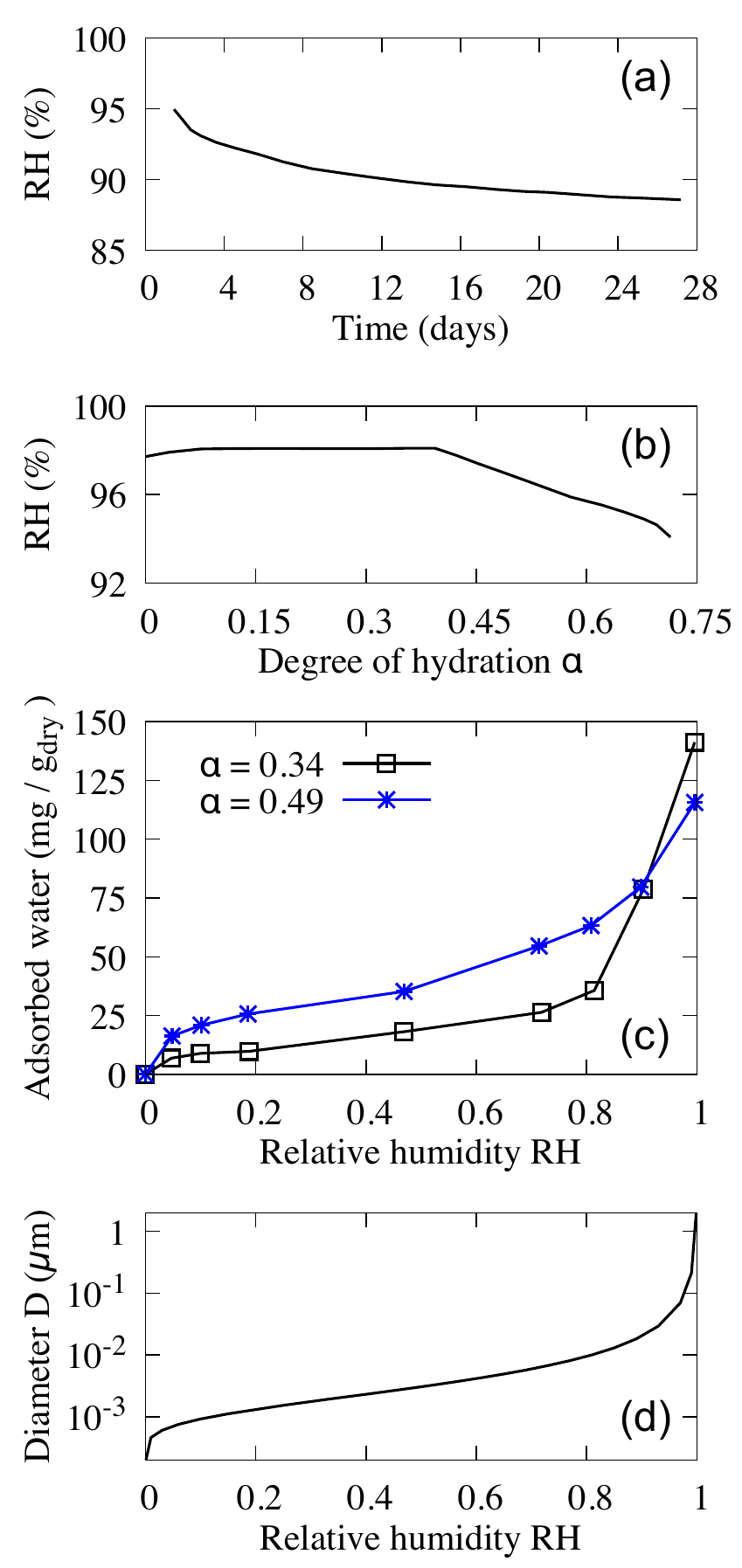}}
\caption{(a) Self-desiccation of a low alkali cement paste (Na$_2$O eq = 0.55\%w), with $w/c=0.3$ and hydrating at 20\Cel (from ref.~\cite{jensen1999influence}); (b) Self-desiccation vs.~degree of hydration $\alpha$ of a low alkali cement paste ($<$0.3\%w) with $w/c=0.35$ and hydrating at 30\Cel (from ref.~\cite{bentz2001influence}); (c) Evolution of the water sorption isotherm during the hydration of cement paste with $w/c=0.2$ {\color{black}at 35\Cel}(from ref.~\cite{Mikhail1972}); (d) Relationship between relative humidity and capillary meniscus diameter, predicted by Kelvin equation {\color{black}at 25\Cel} (see \eqname\ref{eqKelvin}).}
\label{figIntro}
\end{figure}

The hydration of cement paste in sealed conditions involves water consumption and a net loss of volume from reactants to products, known as chemical shrinkage. After setting, chemical shrinkage cannot fully convert to macroscopic shrinkage, thus the chemical activity of water in the paste and the internal relative humidity (iRH) decrease (\eg in \figsname\ref{figIntro}.a and \ref{figIntro}.b). It is generally agreed that the decreasing iRH causes the so-called autogenous shrinkage, although the mechanisms by which pressure and strain are generated are still debated \cite{wittmann2008heresies,ulm2015shrinkage}.

Various models address the autogenous shrinkage of hydrating cement paste and concrete \cite{bazant2000creep,gawin2007modelling,lura2003autogenous,di2009hygro,di2009hygro2}. All these models, at some point, need to relate the hydration-induced consumption of water with the experimentally observed drop of iRH. Usually this is done empirically, either assigning experimental relationships between iRH and degree of hydration $\alpha$, or assigning a water sorption isotherm that relates water content to iRH (water content is relatively easy to predict based on the stoichiometry of the chemical reactions during hydration). The latter approach is more fundamental because the isotherm is a material property that depends on the multiscale pore structure of the paste. The pore structure evolves during hydration, and therefore {\color{black} isotherms are usually presented for different degrees of hydration $\alpha$} (see \figname\ref{figIntro}.c). In principle, sorption isotherms could be predicted from simulated evolutions of the pore structure, which in turn could be related to the mix design of the paste. However sorption isotherms are controlled by the nanopore structure within the hydration product, and current hydration models do not account for such nanopore structure with sufficient detail.

The Kelvin equation relates iRH with the diameter $D$ of the largest water-saturated pore (assuming equilibrium, \ie that all pores smaller than $D$ are saturated, and larger ones are dry):
\begin{eqnarray}
D = - \frac{4\gamma MV_w}{RT \ln RH} \;\;.   \label{eqKelvin} 
 \end{eqnarray} 
\eqname\ref{eqKelvin} assumes perfect wetting and {\color{black}hemispherical liquid-vapour interfaces}. $\gamma = 0.073$ N$\mathrm{m}^{-1}$ is the liquid-vapour surface tension of water, $MV_w = 18.02\cdot10^{-6}$ m$^3$mol$^{-1}$ is the molar volume of water, $R$ is the gas constant, and $T$ is the temperature in Kelvin degrees. At room temperature, $T = 298$ K, the Kelvin equation predicts that iRH starts to decrease appreciably only when sub-micrometre pores get desaturated (see \figname\ref{figIntro}.d): these pores are mostly \textit{within} the hydration product, in particular the porous calcium-silicate-hydrate (C--S--H) gel. {\color{black} The desaturation of C--S--H gel pores requires that the larger capillary pores are desaturated first, and this must occur early during hydration, in order to explain onsets of self-desiccation already during the first day of hydration, as in \figsname\ref{figIntro}.a. At such small degrees of hydration, in pastes with not-very-low water-cement ratio ($w/c\gtrsim0.2$), water consumption by chemical reaction would be insufficient to desaturate the capillary pores. The other possibility is that the C--S--H gel grows rapidly as a low density phase that fills the capillary pores (except for the chemical shrinkage), effectively ``transforming'' them into smaller gel pores.}  To capture early capillary space-filling, some simulations of microstructural development started to assume that the C--S--H gel forms as a very low-density phase that then gets progressively denser with time\cite{Bishnoi2009,konigsberger2016densification}. This assumption is supported by recent results from $^1$H nuclear magnetic resonance (NMR)\cite{Muller201399}, which will be discussed extensively in this manuscript. However, none of the existing hydration models includes details of the pore structure within the gel, thus sorption isotherms and self-desiccation (drop of iRH) are still to be given as empirical constitutive inputs.

Here we combine a simple model of cement hydration, which is focussed on the progressive filling of capillary pores, with a novel description of the evolving pore structure within the C--S--H gel. The latter is informed by recent results from nanoscale simulations of C--S--H gel formation\cite{ioannidou2016mesoscale}. The combined simulations show that a progressive densification, consistent with recent $^1$H NMR results, is necessary in order to  predict the experimentally observed self-desiccation of low-alkali cement pastes. The simulations also show that the presence of dissolved salts in solution alone is not sufficient to explain the observed decrease of iRH. The simulations predict the water sorption isotherms corresponding to the evolving capillary-plus-gel pore size distributions, showing that C--S--H densification is also needed in order to obtain realistic isotherms. Finally, the simulations address the effect of $w/c$ and cement powder fineness on self desiccation and sorption isotherms, and help clarify how the kinetics of C--S--H densification might explain the effect of curing temperature on microstructure and mechanical properties.

\section{Methods}\label{secMethods}

This section presents a simple model for the evolution of pore size distribution in a hydrating cement paste. In some respect, the model is less advanced than state-of-the-art microstructural development simulators \cite{van1991simulation,bentz1994cellular,bullard2007three,bishnoi2009muic,Thomas2011}: (i) it does not track the chemical composition of the solution, (ii) it considers only a statistics of pore sizes, not a 3D configuration of cement grains and hydration products, (iii) it implements only one chemical reaction:
\begin{eqnarray}
\mathrm{C_3S+3.1H \;\; \rightarrow \;\; C_{1.7}SH_{1.8}+1.3CH \;\;,}  \label{eqStoichio} 
 \end{eqnarray} 
where C$_3$S stands for tricalcium silicate, H for water, and CH for calcium hydroxide. The C--S--H in \eqname\ref{eqStoichio} is only the solid part of the gel, hereafter referred to as sCSH as opposed to the C--S--H gel (gCSH) which includes nanopores. Despite its simplicity, the proposed model has two features that are key to describe the evolution of internal humidity and sorption isotherms during hydration: (i) with few seconds of computation it goes from $\alpha = 0$ to 1 describing the evolution of a pore size distribution that spans 6 orders of magnitudes (diameters $D$ from 10$^{-4}$ to 10$^{2}$ $\mu$m, both capillary pores between domains of hydration product and gel pores within the C--S--H); (ii) there is no randomness due to the spatial distribution of phases: alongside the simple chemistry in \eqname\ref{eqStoichio}, this enables a clear analysis of the relationship between C--S--H densification, self-desiccation, and sorption isotherms.
%

{\color{black} Before proceeding with the model description, one should remember that the model presented here is \textit{not} intended to provide a powerful and generalisable chemo-structural description of cement hydration: there are several sophisticated hydration models that already do that in the literature \cite{van1991simulation,bentz1994cellular,bullard2007three,bishnoi2009muic,Thomas2011}. However, none of these sophisticated models is able to capture the self-desiccation and sorption isotherm evolution in a cement paste during early hydration. The model in this paper is \textit{solely} intended to show that what the current sophisticated models are missing is information about the evolution of nanopore structure in the hydration product. If this information is added, even a simple model like the one presented here will predict self-desiccation and sorption isotherms qualitatively well. Following this principle, some components of the model (\eg the evolution of growth and densification rates during hydration) will be kept very simple and specific to the experiments considered here. We prefer to trade some generality and quantitative agreement with the experiments, in exchange for clearer results indicating the importance of modelling nanostructure to describe self-desiccation and isotherms. For a better quantitative agreement, it is suggested that future works should add nanopore-related information to the current state-of-the-art microstructure development models, rather than improving the chemo-structural detail in the simple model presented here.}

\subsection{Model description}\label{secModelDescr}

\begin{figure}
\centerline{\includegraphics[width=0.5\textwidth] {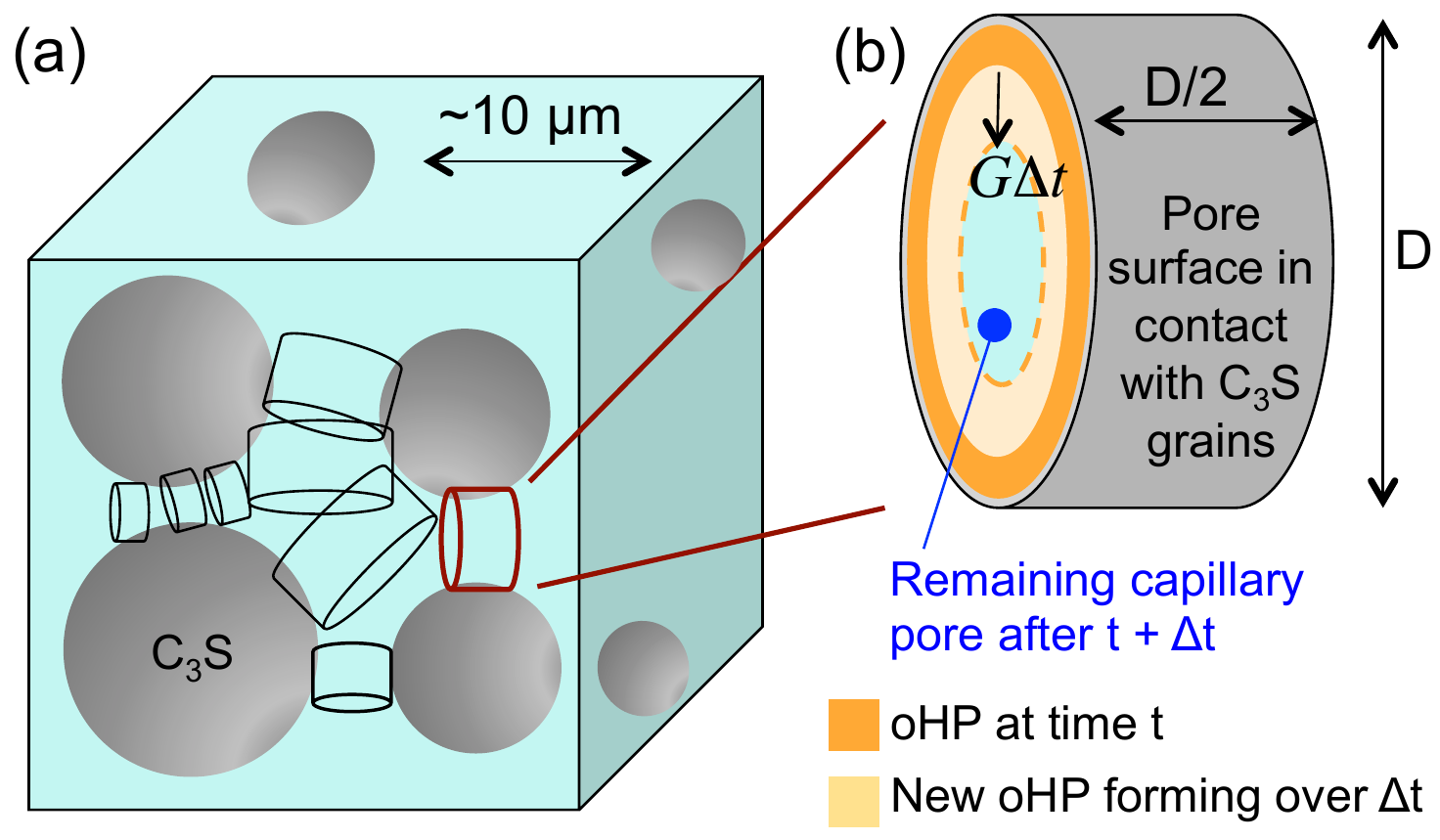}}
\caption{(a) Suspension of C$_3$S grains in water and cylinders discretising the space between grains. (b) Detail of a cylindrical pore at time $t$, displaying the outer product (oHP) already formed in it and new oHP growing radially during the time increment $\Del t$.}
\label{figModel}
\end{figure}

\paragraph{\textbf{Initial pore size distribution (PoSD) at $\mathbf{\alpha =0}$}} For a unit mass of unhydrated cement powder (only C$_3$S  in our model), the initial state of the paste is fully described by the $w/c$ ratio and the specific surface area $S_s$ of the dry powder. At $\alpha = 0$ the capillary pores  can be identified as the spaces between unhydrated cement grains, modelled as a set of cylinders with diameter $D$ and length $D/2$ (see \figname\ref{figModel}.a; the length-to-diameter ratio is arbitrary because it is only affects the number of cylinders $n_p(D)$ introduced below by a constant pre-factor). It is worth pointing out that the model does not consider an actual 3D arrangement of cement grains in suspension: rather, the geometry of the pore network is described only by a pore size distribution $n_{p}(D)$ to be given as an input. $n_{p}(D)$ is the number of capillary pores with diameter $D$ per unit mass of cement powder, or more precisely, $n_p = dN_p(D)/dD$ where $N_p(D)$ is the number of pores with diameter smaller than $D$. The model neglects out-of-equilibrium effects related to the connectivity of the pore network \cite{pinson2014inferring}, hence such connectivity is not described.

It is also assumed that the lateral surface of each cylinder is entirely in contact with cement grains, whereas the circular bases are connected to other capillary pores. Therefore the capillary PoSD must respect two constraints:  
\begin{eqnarray}
& \int_{0}^{\infty} n_{p}(D)V_{1pc}(D)dD = \frac{w/c}{\rho_w}  \;\;,  \label{eqConstrV} \\
& \int_{0}^{\infty} n_{p}(D)S_{1pc}(D)dD = S_s \;\;. \label{eqConstrS}
 \end{eqnarray} 
$V_{1pc}$ and $S_{1pc}$ are the volume and lateral surface of a cylindrical pore with diameter $D$ and length $D/2$. $\rho_w = 1$ g cm$^{-3}$ is the specific weight of water. {\color{black}\eqnames\ref{eqConstrV} and \ref{eqConstrS} are dimensionally homogeneous because $n_p$ is the number of pores (dimensionless) per unit diameter per unit mass of paste.}

In this work, simulations are performed with two types of initial capillary PoSDs {\color{black}, a single-valued pore size distribution, which is the lower bound in terms of pore size variety, and a power-law pore size distribution modelling a fractal agglomeration of cement particles in suspension, hence displaying self-similarity across length scales and allowing in principle for pores of any size. In particular}: 
 \begin{itemize}
 \item For the single-valued distribution: $n_{p}=N\delta_{D\overline D}$, where $\delta$ is the Kronecker delta function. This means that all capillary pores have same initial diameter $\overline D$, representing a perfectly dispersed suspension of cement grains.
 \item For the fractal agglomerate, a distribution with many small pores and few large ones is chosen, $n_{p}= ND^{-\zeta}$, \viz a power law with negative exponent between assigned minimum and maximum pore sizes $D_{min}$ and $D_{max}$ (zero elsewhere). It can be taken $D_{min} = 1$ $\mu$m, which is a typical resolution of microstructure development models, and $D_{max} = 10\overline{D}$ (where $\overline{D}$ is the above-mentioned size if one assumes a single-valued distribution). This represents a flocculated suspension. {\color{black} For all the results in this paper, power law distributions of capillary pores have been created using 2000 linear bins between $D_{min}$  and $D_{max}$}.
\end{itemize}
In both cases the model depends on two unknowns: $N$ and $\overline D$ for the single-valued PoSD, $N$ and $\zeta$ for the power-law PoSD. These unknowns are found \via the constraints in \eqsname\ref{eqConstrV} and \ref{eqConstrS}.

\paragraph{\textbf{Precipitation kinetics}}
It is assumed that C--S--H gel and CH precipitate in two places: (i) in the cylindrical capillary pores, filling them radially from the lateral surface towards the central axis, and (ii) in the new space freed by the dissolution of the C$_3$S (see \eqname\ref{eqStoichio}). 

For precipitation in capillary pores, each pore is treated as an isolated system and it is assumed that the concentration of ions in solution is the same everywhere. \figname\ref{figModel}.b shows a generic capillary pore at a certain hydration time $t$, partly filled with so-called outer hydration product (oHP, which comprises C--S--H and CH). During a time increment $\Del t$, if the pore is still saturated with water, the HP grows {\color{black} at a} rate $G$. The approach to determine whether a pore is saturated or not will be explained later. For a cement paste, $G$ changes in time due to the evolving solution chemistry and morphology of the hydration product at sub-micrometre scales \cite{Thomas2007,bullard2015time,shvab2017precipitation}. {\color{black}To express $G$, one can refer to the Boundary Nucleation and Growth (BNG) model, which is widely used in modelling cement hydration at the sub-micrometre scale \cite{Thomas2007}. According to BNG, the precipitation of HP can be modelled as a set of hemispheres growing radially from initially pointwise nuclei on the surface of the cement grains. The radial growth rate also depends linearly on the supersaturation of the pore solution with respect to HP precipitation, and recent results indicate a $\sim t^{-1}$ decay of such supersaturation during early hydration \cite{bullard2015time}. Mapping these two elements of a BNG model onto the model or radial growth in \figname\ref{figModel}.b leads to:
}
\begin{eqnarray}
G(t) &=& G_{max} \left(\frac{t}{t_{peak}}\right)^2   \;\;\;\; \mathrm{for} \;\;\;\; t \le t_{peak}  \;\;, \label{eqG1} \\
G(t) &=& G_{max} \cdot  \frac{t_{peak}}{t} \;\;\;\; \mathrm{for} \;\;\;\; t > t_{peak}  \;\;. \label{eqG2}
 \end{eqnarray} 
{\color{black}The exponent 2 in \eqname\ref{eqG1} comes from the mechanism of hemispherical growth of HP at the sub-micrometre scale, whereas the $t^{-1}$ decay in \eqname\ref{eqG2} is due to the decrease of solution supersaturation \cite{Thomas2007,scherer2012nucleation,shvab2017precipitation}.} $G_{max}$ and $t_{peak}$ will be calibrated hereinafter with reference to some relevant experiments (\secname\ref{secCalib}). {\color{black} It is worth pointing out that multi-scale mechanics models of cement hydration typically describe the formation of HP using the concept of affinity \cite{ulm1996strength}. In affinity-based models, a differential equation relates the hydration rate with the degree of hydration itself, the latter being considered as an order parameter for the average microstructure of the HP. The BNG model underlying \eqnames\ref{eqG1} and \ref{eqG2} can be considered as the solution of an affinity equation for the specific case of hemispherical HP morphology. In this sense, even if \eqnames\ref{eqG1} and \ref{eqG2} may suggest that $G$ is a direct function of time only, in reality the functional forms of these equations themselves already account for the dependence of hydration rate on HP morphology.}

The oHP is made of CH and C--S--H gel, the latter assumed to form with solid volume fraction $\eta_{min}$ (\viz with internal porosity $= 1-\eta_{min}$):
\begin{eqnarray}
\Del V_{oHP} &=& \Del V_{CHo} + \Del V_{gCSHo} = \Del V_{CHo} + \frac{\Del V_{sCSHo}}{\eta_{min}}  \;\;, \label{eqVout} \\
\frac{\Del V_{CHo}}{MV_{CH}} &= &1.3 \frac{\Del V_{sCSHo}}{MV_{sCSH}} \;\;.  \label{eqMolout} 
 \end{eqnarray} 
\eqname\ref{eqVout} defines the increase of CH and solid C--S--H volumes ($\Del V_{CHo}$ and $\Del V_{sCSHo}$) producing the increase of oHP volume $\Del V_{oHP}$. \eqname\ref{eqMolout} expresses the molar balance from \eqname\ref{eqStoichio} in terms of volume changes.  \tabname\ref{tabMatPar} shows the molar volumes, MV.

\begin{table}[h!]
\centering
\caption{Material parameters used in the simulations.}
\begin{tabular}{c c c c} 
 \hline
 			& Specific weight  			&  Molar volume \\ 
Phase	   	&    $\rho$ (g cm$^{-3}$) 	&  MV  (cm$^3$ mol$^{-1}$)\\ [0.5ex] 
 \hline  \\[-2ex]
 Water 			& 1			&  18.1 \\ 
C$_3$S 			& 3.15	 	&  72.5  \\
Solid C--S--H	& 2.604		&  72.1 \\
CH				& 2.24 		& 33.1   \\
 \hline
\end{tabular}
\label{tabMatPar}
\end{table}

The reactants contributing to the growth of the oHP come from the dissolution of C$_3$S, as per \eqname\ref{eqStoichio}. This dissolution creates new space, which is immediately filled by so-called inner hydration product (iHP). The iHP is made of CH and low-porosity C--S--H gel with solid volume fraction $\eta_{max}$:
\begin{eqnarray}
\Del V_{iHP} &=& \Del V_{CHi} + \Del V_{gCSHi} = \Del V_{CHi} + \frac{\Del V_{sCSHi}}{\eta_{max}}  \;\;, \label{eqVin} \\
\frac{\Del V_{CHi}}{MV_{CH}} &= &1.3 \frac{\Del V_{sCSHi}}{MV_{sCSH}} \;\;.  \label{eqMolin} 
 \end{eqnarray} 
 
The formation of iHP requires reactants from the C$_3$S too. Therefore the volume and moles of dissolved C$_3$S must be balanced by the volumes of precipitated iHP and by the moles of precipitated sCSH (\eqname\ref{eqStoichio} shows that each mole of dissolved C$_3$S corresponds to one mole of precipitated C--S--H):
\begin{eqnarray}
\Del V_{C_3S} &=& \Del V_{iHP}  \;\;, \label{eqVC3S} \\
\frac{\Del V_{C_3S}}{MV_{C_3S}} &= & \frac{\Del V_{sCSH}}{MV_{sCSH}} =  \frac{\Del V_{sCSHo}+\Del V_{sCSHi}}{MV_{sCSH}} \;\;.  \label{eqMolC3S} 
 \end{eqnarray} 
 
 \eqsname\ref{eqVout}-\ref{eqMolC3S} form a system of 6 equations in 6 unknowns: all the $\Delta V$'s except for $\Del V_{oHP}$ which comes directly from $G\Del t$.  In this way, for each $\Del t$ and for each cylindrical capillary pore, one can compute the volume and mole increments for all the solid reactants and products. The role of water will be discussed later. {\color{black}. The time step increases logarithmically during the simulations in this paper, starting from $\Del t =0.01$ days when $t=0$ and increasing until $t = 456$ days, \viz \ca 15 months.}

\paragraph{\textbf{C--S--H gel densification}}
After each $\Del t$, the volume of newly formed C--S--H gel in the oHP is recorded {\color{black} as last entry in} a vector \textbf{$\mathbf{\Del V_{gCSHo}}(t)$}, where $t$ is the time at which the generic  {\color{black} element of the vector} $\Del V_{gCSHo}(t)$ has formed. The basic hypothesis put forward in this work is that each $\Del V_{gCSHo}(t)$ gets progressively denser with time, as long as its gel pores are saturated with water. To model the densification of the gel, the solid volume fraction $\eta$ of each $\Del V_{gCSHo}$ increases with time from $\eta_{min}$ to $\eta_{max}$ (same as the $\eta_{max}$ assigned immediately to the C--S--H gel in the iHP, as per \eqname\ref{eqVin}) with rate:
\begin{eqnarray}
\dot\eta(t)  &=& 0  \;\;\;\; \mathrm{for} \;\;\;\; t \le t_{t_{d0}}  \;\;, \label{eqDenst1} \\
\dot\eta(t)  &=& k \left(\frac{t_{ch\eta}}{t-t_{d0}+t_{ch\eta}} \right)  \;\;\;\; \mathrm{for} \;\;\;\; t > t_{t_{d0}}  \;\;. \label{eqDenst2}
 \end{eqnarray} 
\eqnames\ref{eqDenst1} and \ref{eqDenst2} assume that the densification starts only after a time $t_{d0}$ with rate $k$, and then progressively decelerates with a characteristic time scale $t_{ch\eta}$. Both the deceleration of oHP growth (decreasing $G$ in \eqname\ref{eqG2}) and the deceleration of C--S--H densification are driven by depletion of reactants in solution \cite{bullard2015time,shvab2017precipitation}, and therefore it is reasonable to assign the same characteristic timescale to both processes: $t_{ch\eta}=t_{peak}$. The parameters $k$ and $t_{d0}$ will be calibrated in \secname\ref{secCalib}. 

{\color{black}The proposed formulation for $\dot\eta$ entails two rate discontinuities: a significant one at time $t_{d0}$, when $\dot\eta$ jumps from 0 to $k$, and one at $\eta = \eta_{max}$, when $\dot\eta$ jumps back to 0. For the case studies in this paper, the latter discontinuity is negligible because, by then, $\dot\eta$ has already decreased almost to zero due to its $t^{-1}$ evolution. Regarding the discontinutity at $t_{d0}$, one could remove it by making the realistic assumption that densification starts immediately, \viz $t_{d0}=0$. However, that would cause the hydration rate $\dot{\alpha}$ at early age, \ie when the growth rate $G$ from \eqnames\ref{eqG1} and \ref{eqG2} is still relatively large, to depend on both $G_{max}$ and $k$. This would complicate the calibration of these rate constants, whereas taking  $t_{d0}$ sufficiently greater than $t_{peak}$ will allow calibrarting $G_{max}$ and $k$ independently and, in part, analytically (see \secname\ref{secCalib}). Furthermore, turning off densification during the first hours and letting it start only when the oHP growth is slow, will lead to a clear separation of regimes: from growth-controlled hydration to densification-controlled hydration. In this way it will be straightforward to associate the computed evolution of self-desiccation and sorption isotherms with the underlaying rate-controlling mechanism.}

{\color{black} Recent analyses of experimental data, by K\"osnigsberger et al.~\cite{konigsberger2016densification}, indicate that the density of the C--S--H gel averaged over a whole macroscopic sample of cement paste, is a function of the water-to-cement ratio ($w/c$) of the paste itself. This may either imply that $k$ in \eqnames\ref{eqDenst2} is a direct function of $w/c$, \viz that pointwise densification depends on macroscopic confinement possibly due to ion diffusion effects, or that that dependence on $w/c$ is only true on average and results from the process of filling capillary pores with various different sizes. Not having theoretical nor experimental data on the relationship between local densification and confinement, this paper will assume that $k$ is independent of $w/c$ and test the latter of the two hypotheses above in the Results section.}

{\color{black}The $t^{-1}$ scaling in \eqname\ref{eqDenst2} implies that $\dot{\eta}(t)$ scales as $\exp[\eta(t_{d0}) - \eta(t)]$. This differential equation is analogous to affinity-based models of hydration rate in multiscale concrete mechanics (there the degree of hydration $\alpha$ is considered instead of the gel density $\eta$, but the two are related by stoichiometry). Such an exponential relationship between $\dot{\eta}$ and ${\eta}$ may reflect the exponential decay of interconnected paths (here paths for water and ion diffusion) in a random percolation model of densification \cite{duminil2017exponential}. Therefore, even if \eqname\ref{eqDenst2} may suggest that densification is only a function of the material's age, actually the functional form of \eqname\ref{eqDenst2} implies a specific assumption regarding the morphology of the hydration product (here, for example, a densifying random agglomeration blocking diffusion paths). In this work, the $\dot{\eta}(t) \sim \exp[- \eta(t)]$ scaling was actually chosen in order to recover a $t^{-1}$ scaling in \eqname\ref{eqDenst2}. The reason for this is heuristic, because experimental measurements show that the degree of hydration $\alpha$ evolves logarithmically over years of hydration, because the $\dot{\eta}\sim t^{-1}$ scaling in \eqname\ref{eqDenst2} leads indeed to $\eta\sim \log(t/t_{ch\eta})$, and because $\alpha$ scales as $\eta$ in our model when hydration starts to be controlled by C--S--H gel densification (\ca after the first day of hydration, as shown in the Results section). 
}

In terms of volume and mass balance, densification implies the precipitation of additional volumes of HP, $\Del V_{HPd}$, made of CH and solid C--S--H, $\Del V_{CHd}$ and $\Del V_{sCSHd}$. Both these volumes are assumed to fill the pores of the densifying C--S--H gel in the oHP. This brings the advantage of uncoupling densification from capillary space filling. It is worth observing that other scenarios are possible but are not explored here, \eg one could assume that only $\Del V_{sCSHd}$ determines densification whereas $\Del V_{CHd}$ precipitates in the capillary pores. 

The additional HP volumes change \eqname\ref{eqMolC3S} to:
\begin{eqnarray}
\frac{\Del V_{C_3S}}{MV_{C_3S}} &= & \frac{\Del V_{sCSH}}{MV_{sCSH}} =  \frac{\Del V_{sCSHo}+\Del V_{sCSHi} + \Del V_{sCSHd}}{MV_{sCSH}} \;\;.  \label{eqMolC3Sd} 
 \end{eqnarray} 
Furthermore:
 \begin{eqnarray}
\Del V_{HPd} &=& \Del V_{CHd} + \Del V_{sCSHd} \;\;, \label{}  \\
\frac{\Del V_{CHd}}{MV_{CH}} &= &1.3 \frac{\Del V_{sCSHd}}{MV_{sCSH}} \;\;.  \label{eqMold} 
 \end{eqnarray} 
$\Del V_{HPd}$ is computed by integration of \eqnames\ref{eqDenst1} and \ref{eqDenst2}, thus \eqnames\ref{eqVout}-\ref{eqVC3S},\ref{eqMolC3Sd}-\ref{eqMold} form a system of 8 linear equations in 8 unknowns.

\paragraph{\textbf{Pore size distribution (PoSD): capillary and gel pores}} At each time, the PoSD of the capillary pores is obtained by subtracting the thickness of the oHP layer inside each cylinder from the original cylinder size (see \figname\ref{figModel}.b), if one neglects the small contraction due to autogenous shrinkage. The PoSD inside the C--S--H gel instead must be given as a constitutive input. Simulations based on the aggregation of nano-units of solid C--S--H provide 3D models of the internal structure of the C--S--H gel \cite{Masoero2012,gonzalez2013nanoscale,del2014soft,etzold2014growth,ioannidou2016mesoscale}, and are also starting to link gel morphology with solution chemistry \cite{shvab2017precipitation}. The PoSD of two such simulated structures with different solid volume fraction $\eta$ have been published recently \cite{ioannidou2016mesoscale}. They indicate that the gel is made of two domains: a dense domain $\delta$ of highly aggregated nano-units with small pores (average diameter below 2 nm), and a loose domain $\lambda$ with bigger pores. The {\color{black} same nanoscale simulations in Ref.~\cite{ioannidou2016mesoscale}} also show that, as $\eta$ increases: (i) the dense $\delta$-domain occupies a progressively large fraction of the gel volume, hence the gel pores in the dense domain occupy a larger fraction $f_\del$ of the gel pore 
\begin{figure}[t]
\centerline{\includegraphics[width=1\textwidth] {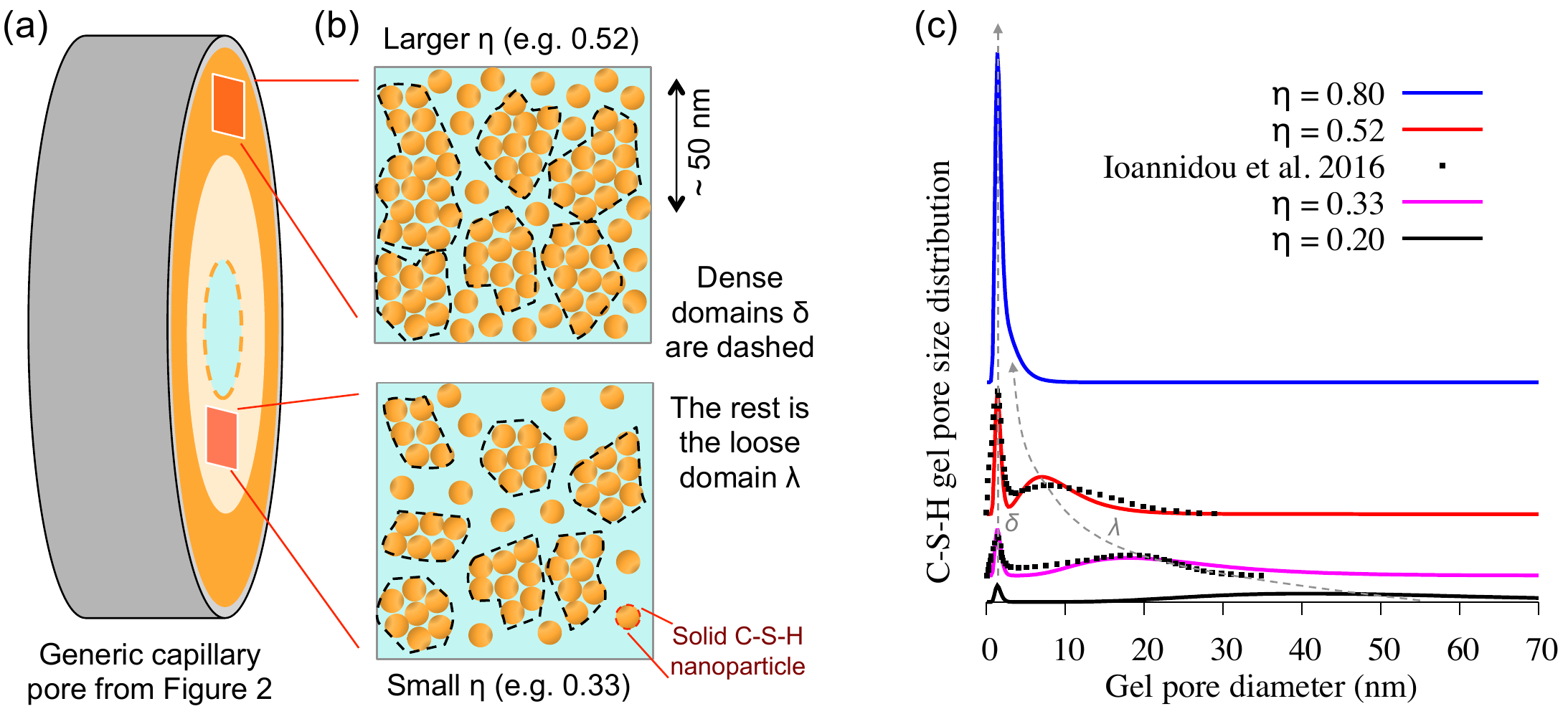}}
\caption{{\color{black}Schematic evolution of C--S--H morphology and nano-pore structure with solid volume fraction $\eta$. (a) A cylindrical capillary pore partially filled by hydration product with a radial gradient of $\eta$, because $\eta$ depends on when the product precipitated during hydration (\cf \figname\ref{figModel}). (b) Based on nanoparticle simulation results from Ref.~\cite{ioannidou2016mesoscale}, the hydration product is assumed to always host a local coexistence of dense $\delta$ and loose $\lambda$ domains. An increase of $\eta$ induces two effects: first, an increase of volume occupied by dense domains at the expense of loose ones; second, an increase of local solid fraction within the loose domains, whereas the local solid fraction within dense domains does not change with $\eta$. (c) Gel PoSDs reflecting the morphology changes in (b) as $\eta$ increases. The PoSDs are obtained by fitting and extrapolating nanoparticle simulation results from Ref.~\cite{ioannidou2016mesoscale}.}}
\label{figDens}
\end{figure}
volume, (ii) some of the gel pores in the dense domain disappear getting filled with solid, but (iii) the PoSD within the dense domain does not change significantly. All this is shown in \figname\ref{figDens}.a. Meanwhile the pores in the loose domain occupies a progressively smaller volume fraction of the gel porosity $f_\lambda$ (because $f_\del+f_\lambda = 1$ always), with their PoSD displaying a progressively smaller average pore size (from \ca 20 nm when $\eta = 0.33$ to \ca 8 nm when $\eta = 0.52$ \cite{ioannidou2016mesoscale}). This mechanism describes gel densification as progressive filling of gel pores with newly precipitated solid: other mechanisms involving the ageing of already formed solid C--S--H \cite{bauchy2015creep}, \eg due to polimerization of silicate chains \cite{di2009hygro}, are not considered here as their effect on the overall gel density is likely to be comparatively much smaller.

To capture the nanoscale simulation results, one may describe the gel PoSD as the sum of two log-normal distributions, one for the pores in the $\delta$ domain and one for those in the $\lambda$ domain:
{\color{black}
\begin{eqnarray}
\frac{1}{\eta V_{gCSH}} \frac{dV_{pg}(D)}{dD} &= & \frac{1}{\eta V_{gCSH}} \left( \frac{dV_{pg}^\delta(D)}{dD}  + \frac{dV_{pg}^\lambda(D)}{dD}  \right)= \nonumber \\
&=& \frac{1}{D\sqrt{2\pi}}   \left\{ \frac{f_\del}{\frac{\sigma_\del}{D_0}}\exp\left[ - \frac{D_0^2 \left(\ln \frac{D}{D_0} - \mu_\del \right)^2}{2\sigma_\del^2}\right]   +    \frac{f_\lambda}{\frac{\sigma_\lambda}{D_0}}\exp\left[ - \frac{D_0^2(\ln \frac{D}{D_0} - \mu_\lambda)^2}{2\sigma_\lambda^2}\right]  \right\}    \;\;.  \label{eqPoSDg} 
 \end{eqnarray} 
$V_{gCSH}$ is a generic volume of C--S--H gel, be it part of the inner or outer product, and formed at any time; thus $\eta V_{gCSH}$ is the gel pore volume at a generic location inside the paste. $V_{pg}(D)$ is the cumulative volume of gel pores with diameter below $D$, thus $f_\del = (\eta V_{gCSH})^{-1}\int_{D=0}^\infty {dV_{pg}^\delta(D)}{dD}$, and $f_\lambda= (\eta V_{gCSH})^{-1}\int_{D=0}^\infty {dV_{pg}^\lambda(D)}{dD}$. The arbitrary scale factor $D_0$ is set to 1 $\mu$m.The $\mu$, $\sigma$, and $f_\del$ parameters of the distribution are fitted to match the gel PoSDs from nanoscale simulations, as shown in \figname\ref{figDens}.b ($f_\lambda$  then comes from $f_\del+f_\lambda = 1$):
\begin{eqnarray}
f_\del &=& \eta^{2.1} \;\;, \label{eqFd} \\
D_0\; e^{\mu_\del} &=& 0.0015  \;\; \mathrm{\mu m}\;\;, \label{eqMd}\\
\sigma_\del &=& 0.25  \;\; \mathrm{\mu m}  \;\;, \label{eqSd}\\
D_0 \; e^{\mu_\lambda} &=& 0.0015^{(\eta^{0.48})}  \;\; \mathrm{\mu m}\;\;, \\
\sigma_\lambda &=& 0.45  \;\; \mathrm{\mu m} \;\;. \label{eqSl}
 \end{eqnarray} 
 } 
 \eqname\ref{eqFd} implies that the volume fraction of pores in the dense gel domain $f_\del$ increases with the gel solid fraction $\eta$, tending to 1 as $\eta \rightarrow 1$. \eqsname\ref{eqMd} and \ref{eqSd} imply that the PoSD of the dense domain does not depend on $\eta$ and has a median pore size $e^{\mu_\del} = 1.5$ nm. The loose domain instead has median pore size $e^{\mu_\lambda}$ which decreases with $\eta$ tending to $e^{\mu_\del}$ as $\eta \rightarrow 1$. When $\eta$ tends to 0,  \eqnames\ref{eqFd}-\ref{eqSl} imply that $f_\del\rightarrow 0$, thus $f_\lambda \rightarrow 1$ (the whole gel is loose) with median pore size $e^{\mu_\lambda} \rightarrow 1$ $\mu$m: in such limit, gel pores are not discernible from capillary spaces between C$_3$S grains.

{\color{black} In the simulations in this paper, gel pore size distributions are described using 20,000 linear bins between minimum and maximum diameters of 0.1 nm and 2 $\mu$m. Total PoSDs, including capillary and gel pores, are instead computed using 100 logarithmic bins between minimum and maximum diameters of 0.1 nm and $D_{max}$, the latter being the maximum diameter of capillary pores at time $t=0$.}

\paragraph{\textbf{Water consumption, saturation, and internal relative humidity}}
Knowing the volume of C$_3$S consumed at each $\Delta t$, the corresponding volume of reacted water $\Del V_w$ is (see \eqname\ref{eqStoichio}):
\begin{eqnarray}
\Del V_w &=& 3.1 \frac{MV_w}{MV_{C_3S}}\Del V_{C_3S} \;\;. \label{eqDVw}
 \end{eqnarray} 
If sealed hydration is considered, $\Del V_w$ in \eqname\ref{eqDVw} is the only contribution to changes in water content. The volume of water $V_w(t)$ that is still present at time $t$ partially saturates the gel and capillary pores (not entirely, because of chemical shrinkage). The location of $V_w(t)$ determines the internal relative humidity iRH of the sample, controlled by the curvature of the water liquid-vapour interface (capillary meniscus) as per Kelvin equation (\eqname\ref{eqKelvin}). The location of $V_w(t)$ also determines where new hydration product can precipitate, because the oHP can only grow in saturated capillary pores and only saturated C--S--H gel in the oHP can densify.

To compute the iRH one can assume that water is in its equilibrium distribution, \ie preferentially saturating the smallest pores first. This means that if the cumulative volume of all pores (gel plus capillary) with diameter smaller than a certain $D_{men}$ equals $V_w(t)$, then the water liquid-vapour interface will sit in pores with diameter $D_{men}$ and the capillary meniscus will have diameter $D_{men}$ too. At the generic time $t$ the proposed model provides the gel and capillary PoSD, thus one only needs to identify the pore size $D_{men}$ such that: 
\begin{eqnarray}
\int_{D=0}^{D_{men}} \frac{dV_p(D)}{dD} dD = V_w(t)\;\;. \label{eqDmen}
 \end{eqnarray} 
$V_p$ is the volume of all pores, gel and capillary, with diameter smaller than $D$. Knowing $D_{men}$, \eqname\ref{eqKelvin} provides the iRH. It is important to point out that this approach neglects the water adsorbed on the surface of otherwise dry pores, because: (i) the amount of such water is usually small compared to the sum of water adsorbed \via capillary condensation and water in the interlayer spaces of solid C--S--H \cite{pinson2015hysteresis} (the latter is accounted for in \eqname\ref{eqStoichio}); (ii) the impact of surface water on the diameter of the capillary meniscus is important only in small pores, \viz at low RH, whereas the focus of the present study is on on self-desiccation and on the shape of the sorption isotherm at large RH.

To determine which capillary pores are saturated, and therefore where new HP can and cannot grow, it is assumed that all capillary pores with $D\le D_{men}$ are saturated. For the gel pores the same approach could be used, but that would require a gel densification law defined on an individual gel pore basis, whereas the densification rate in \eqname\ref{eqDenst2} is assigned to the whole C--S--H gel volume formed at a certain time, and to all the pores within it. Therefore, in order to decide whether a gel volume is saturated or not, the total volume of gel pores in the sample is compared to $V_w(t)$. If it is smaller, then all gel pores are saturated and densification can proceed everywhere. If larger, the model assumes that the first gel volumes to dry up and stop densifying are the ``youngest'' ones, \viz those that formed at later times. The rationale for this is that $\eta$ increases with time and that the median gel pore size decreases with $\eta$, hence large gel pores will prevail in ``young'' C--S--H gel. In this way one can identify a formation time $\tau_{sat}$ of the C--S--H gel such that the gel $\Delta V_{gCSHo}$ that formed at $\tau>\tau_{sat}$ is dry whereas the gel that formed before is saturated:
\begin{eqnarray}
\int_{\tau=0}^{\tau_{sat}} \left[1-\eta(t,\tau)\right] \Delta V_{gCSHo}(\tau) d\tau = V_w(t) \;\;. \label{}
 \end{eqnarray} 
$\eta(t,\tau)$ is the solid volume fraction at time $t$ of a C--S--H gel volume formed at time $\tau$.

\begin{figure}
\centerline{\includegraphics[width=0.5\textwidth] {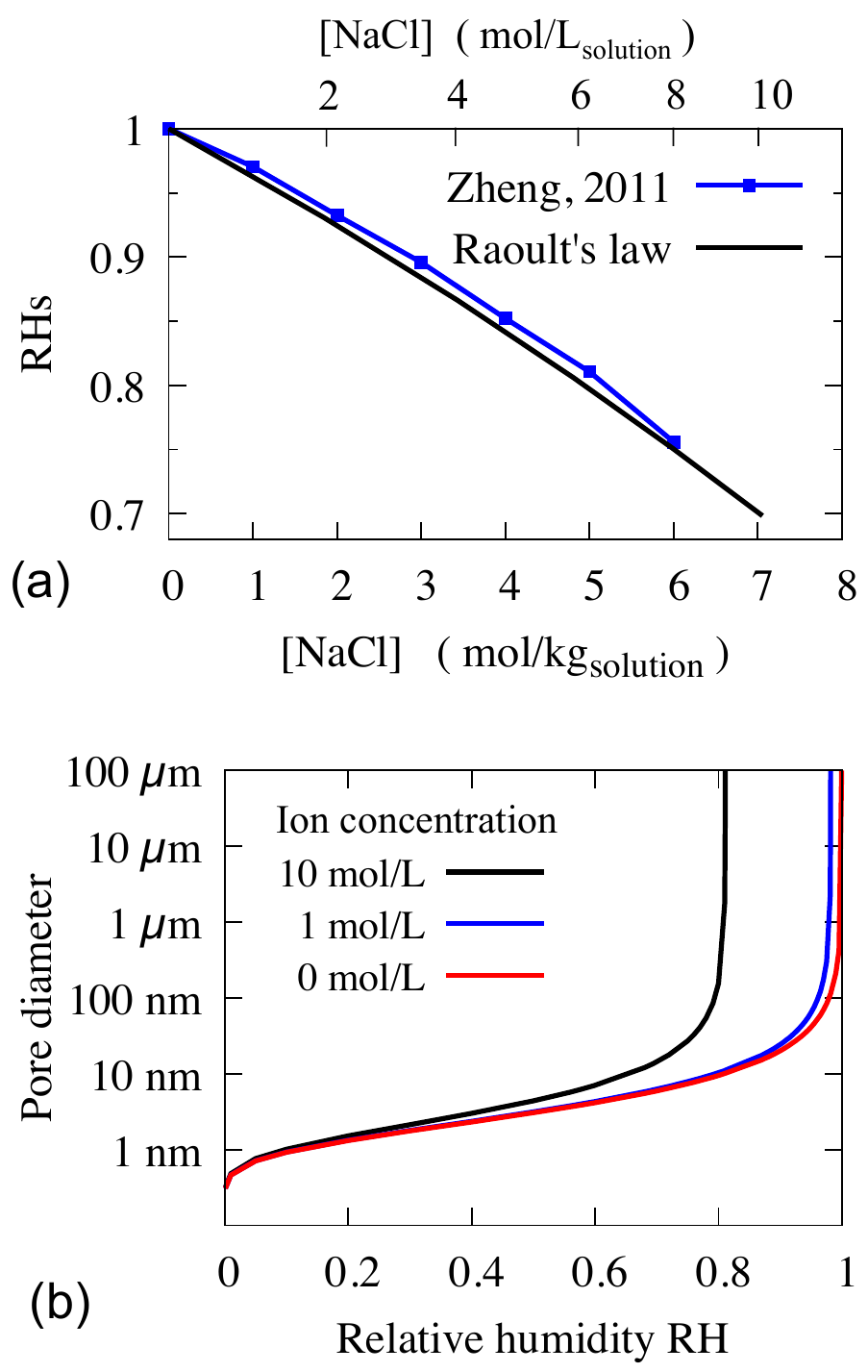}}
\caption{(a) Effect of ions in solution on the saturation relative humidity of water, RHs. In our calculations (Raoult's law) we considered molar concentration of liquid water = 55.5 mol L$^{-1}$, apparent molar volume of NaCl in aqueous solution = 16.6 cm$^3$ mol$^{-1}$, molar mass of NaCl = 58.44 g mol$^{-1}$,  and that the NaCl dissociates in solution.  (b) Effect of ions in solution on the diameter of the largest saturated pore, as per modified Kelvin equation in \eqname\ref{eqKelvinS}. We considered ions with apparent molar volume of 20 cm$^3$ mol$^{-1}$, which is a large value for typical ions in a cement solution\cite{marcus2009standard} (larger molar volumes affect more the molar fraction of water $X_w$ and RHs).
}
\label{figNaCl}
\end{figure}

\paragraph{\textbf{Dissolved salts}}
The liquid water in a hydrating cement paste is actually an ionic solution which can reach concentrations of several moles per litre due to dissolution of salts. The ions in solution reduce the saturation relative humidity RHs of water below 1, modifying the Kelvin equation \cite{lura2003autogenousTH}:
\begin{eqnarray}
D = - \frac{4\gamma MV_w}{RT \ln \frac{RH}{RHs}} \;\;.   \label{eqKelvinS} 
 \end{eqnarray} 
The easiest way to estimate RHs is to consider the solution as an ideal mixture and invoking Raoult's law to obtain RHs $= X_w$ (the molar fraction of liquid water in the solution) \cite{lura2003autogenousTH}. For a solution of NaCl in water, \figname\ref{figNaCl}.a shows that the RHs predicted by the ideal mixture model agrees with predictions from more sophisticated approaches analysed in Ref.\cite{zeng2011poromechanical}.  \figname\ref{figNaCl}.b shows the effect of this simple correction on the Kelvin equation.

In a cement solution, the most concentrated species are alkalis (calcium ions and silicates are negligible because their concentrations are capped to millimolar values by the  low solubility of C--S--H and CH). The mass of alkalis in a dry cement paste is often given as equivalent sodium dioxide, Na$_2$Oeq (the molar volume of Na$_2$O is 62 g mol$^{-1}$). The calculations in this paper we will always assume that all the alkalis get immediately dissolved, hence maximising the impact of dissolved salts on RHs. The simulations will track the moles of unreacted water, hence computing $X_w$ = RHs will be straightforward. It is worth noting that the apparent volume of ions in solution may cause the volume of solution to differ, very slightly, from the volume of pure unreacted water. This effect is neglected in this study.

\paragraph{\textbf{Water sorption isotherms}} Knowing the volume and size distribution of gel and capillary pores during hydration, the simulations provide water sorption isotherms that evolve with time and degree of hydration $\alpha$. The isotherm {\color{black} is obtained by summing together two contributions of adsorbed water, which are computed independently for each level of relative humidity}: water in the interlayer spaces of the solid C--S--H, and water condensed in the gel and capillary pores. The water adsorbed on the surface of otherwise dry pores is neglected, for the reasons discussed above, after \eqname\ref{eqDmen}. 

\begin{figure}[h]
\centerline{\includegraphics[width=0.5\textwidth] {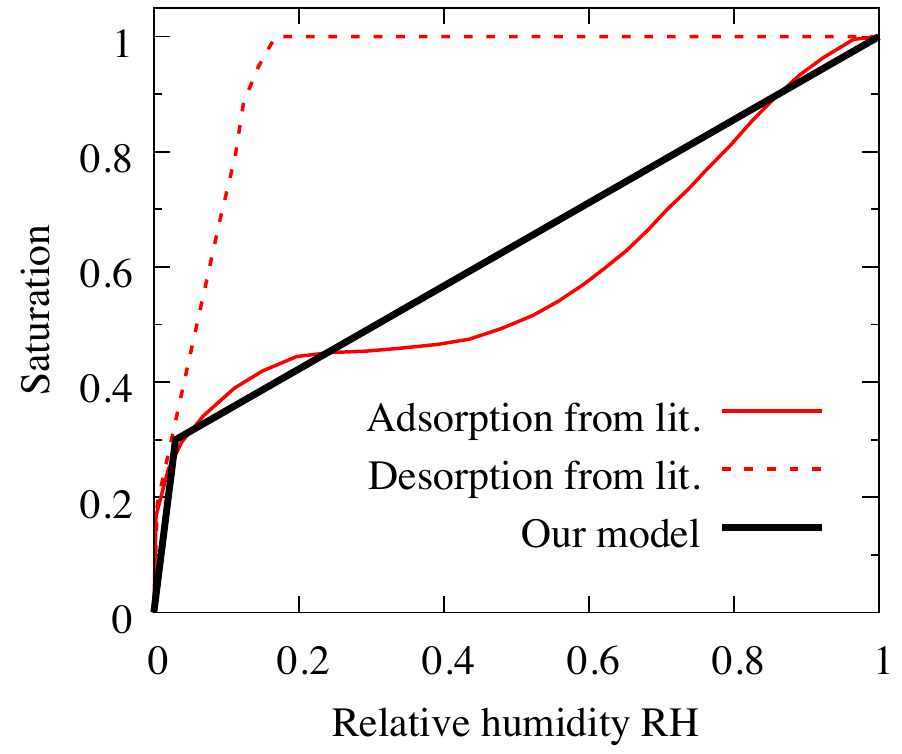}}
\caption{Sorption isotherm for interlayer water in solid C--S--H.  {\color{black} Literature results refer to experiments by Feldman \cite{Feldman1968s} recently reviewed by Pinson et al.~\cite{pinson2015hysteresis}.}}
\label{figIsoInter}
\end{figure}

For the total amount of interlayer water at time $t$ the model considers all the 1.8 moles of water in the solid C--S--H per mole of reacted C$_3$S in \eqname\ref{eqStoichio}. This corresponds to the $\frac{1.8}{3.1}= 54.4 \%$ of all the water that has reacted up to time $t$. This is an upper bound, because some of the water in the solid C--S--H is actually chemically bound as OH groups (see \eg Ref.\cite{manzano2012confined}). Furthermore, the model does not consider that the amount of water going into the solid C--S--H when it forms (\eqname\ref{eqStoichio}) should be a function of the evolving iRH: this is a reasonable approximation because the iRH is unlikely to decrease significantly below 90\% during the self desiccation of low-alkali pastes considered here. Going back to the isotherm, {\color{black} experiments by Feldman \cite{Feldman1968s}, recently reviewed in relation to molecular simulation results in Ref.~\cite{pinson2015hysteresis}, indicate that the interlayer water follows the adsorption and desorption isotherms shown in \figname\ref{figIsoInter}. In the present paper, only adsorption isotherms will be computed, thus only the adsorption branch in \figname\ref{figIsoInter} is considered. Furthermore, in order to simplify the computation, the adsorption branch is approximated using a bilinear function, also shown in \figname\ref{figIsoInter}, which is based on the observations} that the interlayer spaces are fully saturated when RH = 1, and 30\% saturated when RH = 0.03. {\color{black} This is sufficient in order to estimate the part of the total sorption isotherm which depends on interlayer water}.

{\color{black} The rest of the total sorption isotherm depends on water in gel and capillary pores, which are larger than the interlayer spaces and are controlled by} capillary condensation. {\color{black} The sorption isotherm for these larger pores is computed following the steps below}:
\begin{enumerate}
\item Fix a relative humidity RH; 
\item Use the Kelvin equation in \eqname\ref{eqKelvin} to calculate the diameter of the capillary meniscus $D_{men}$ which is the largest saturated pore. For simplicity this step neglects the effect of dissolved salts, which would require an iterative process to find the $D_{men}$ that is consistent with the RH$_s$ reduced by the concentration of ions in solution, with this latter concentration depending on the adsorbed water and thus on $D_{men}$. The results will show that, in low-alkali pastes, ions in solution, have a limited importance and therefore their impact on the isotherms would also be small, except maybe at very low RH when the ions get more concentrated due to scarcity of adsorbed water. The discussion of simulated sorption isotherms will focus instead on quite large RH $> 0.5$;
\item Use the gel and capillary PoSDs at time $t$ from the hydration model to compute the total volume of pores with $D\le D_{men}$, which give the total water content and thus the saturation;
\item Repeat for different values of RH between 0 and 1. 
\end{enumerate}
These steps provide the total amount of water that would be absorbed by capillary condensation if the hydration of the paste stopped  at time $t$ and if the paste were brought to equilibrium with a given environmental RH. This approach neglects out-of-equilibrium water distributions, which arise especially during drying (\eg the ``ink--bottle'' effect): these would be important in order to predict hysteresis in the isotherms \cite{pinson2015hysteresis}. Assuming equilibrium however leads to a simpler model and will be sufficient to support the analysis of the presented results.

The last step to compute the isotherm is to normalise the mass of adsorbed water by the dry mass of the paste $m_{dry}$. The dry mass increases during hydration, because it comprises the initial mass of binder plus the water that gets chemically bound during hydration. This chemically bound water is identified with the water that produces CH in \eqname\ref{eqStoichio}, which is consistent with the previous assumption that all the water in the interlayer space of the solid C--S--H is evapourable.

\subsection{Model calibration: hydration and densification kinetics}\label{secCalib}
Six parameters need calibration: the minimum and maximum possible solid fraction of the C--S--H gel ($\eta_{min}$ and $\eta_{max}$), the maximum growth rate ($G_{max}$), the characteristic time scale of growth and densification ($t_{peak}$), the maximum {\color{black} (\ie initial)} densification rate ($k$), and the time at which densification starts ($t_{d0}$). 

$\eta_{max}$ can be taken directly from the available literature, which indicates $\eta_{max}=0.74$ as the solid volume fraction of so-called high-density C--S--H gel \cite{jennings2000model}. This parameter controls the asymptotic pore size distribution and thus the sorption isotherm of pastes with low $w/c$ hydrated underwater, whose asymptotic $\alpha$ is determined by space filling and not by the availability of reactants. $\eta_{max}$ also determines the solid fraction of the iHP, which here is assumed to form immediately. Increasing $\eta_{max}$ would thus increase the hydration rate $\Delta  \alpha/\Delta t$: if this rate is the experimental quantity to be captured, a larger $\eta_{max}$ could be compensated by smaller growth and densification rate constants $G_{max}$ and $k$.

In the presented model, the C--S--H gel in the oHP starts to densify only after hydrating for a time $t_{d0}$. One can assume $t_{d0}=1$ day and $t_{peak} = 10$ hours, the latter being the time at which the growth rate $G$ reaches its maximum value before starting to decelerate (a peak of early hydration rate at 10 hours is typical in ordinary cement pastes \cite{Bullard2011}). In this way early hydration, controlled by HP growth at $t<t_{d0}$, is uncoupled from later hydration occurring at $t>t_{d0}$, when $G$ has already significantly decreased and hydration is controlled by the rate of gel densification. In real pastes growth and densification coexist to some extent, especially during early hydration. The presented parametrization, however, simplifies the calibration and the interpretation of the results.

$\eta_{min}$ controls the density of just-formed C--S--H gel in the outer hydration product. In particular, it controls the density of \textit{all} the outer C--S--H gel during early hydration, when $t<t_{d0}$. A number of  experiments display self-desiccation already during early hydration, thus for such experiments $\eta_{min}$ must be sufficiently small to ensure that all the capillary pores are almost filled by oHP when $t<t_{d0}$ (see \secname\ref{secResults} for discussion of this point). One such experiment, showing early self-desiccation at  $\alpha = 0.4$ and $t<1$ day, is a low-alkali cement paste studied by Bentz et al.~\cite{bentz2001influence}, with $w/c=0.35$ and Blaine fineness of the dry powder $S_s = 387$ m$^2$ kg$^{-1}$. A scenario with all capillary pores filled at $t<t_{d0}$ leads to the following equations, where all the $\alpha$-dependent quantities refer to $\alpha = 0.4$ for the experiment by Bentz et al.:
\begin{eqnarray}
V_{gCSHi}(\alpha) + V_{CHi}(\alpha) &=& \alpha   V_{C3S,0}    \;\;, \label{eqVinCal} \\
V_{gCSHo}(\alpha) + V_{CHo}(\alpha) &=&   V_{w,0}    \;\;, \label{eqVoutCal} \\
\frac{V_{sCSHi}(\alpha)+V_{sCSHo}(\alpha)}{MV_{sCSH}} &=& \frac{\alpha V_{C3S,0}}{MV_{C3S}}   \;\;. \label{eqMolCal}
 \end{eqnarray} 
\eqname\ref{eqVinCal} states that the volume of dissolved C$_3$S at $\alpha = 0.4$ (with $V_{C3S,0}$ being the initial volume of C$_3$S) is filled by inner C--S--H gel and CH. \eqname\ref{eqVoutCal} states that the volume initially occupied by water, $V_{w,0}$, must be filled by outer C--S--H gel and CH when self-desiccation starts. \eqname\ref{eqMolCal} is the molar balance between dissolved C$_3$S and solid C--S--H. By setting $V_{gCSHi}= V_{sCSHi}/\eta_{max}$ and $V_{gCSHo} = V_{sCSHo}/\eta_{min}$, and by invoking \eqnames\ref{eqMolout} and \ref{eqMolin}, one obtains a system of 5 linear equations in 5 unknowns, providing $\eta_{min} = 0.195$. A smaller value of  $\eta_{min}$ would lead to self-desiccation at even smaller $\alpha$. 

\begin{figure}
\centerline{\includegraphics[width=1\textwidth] {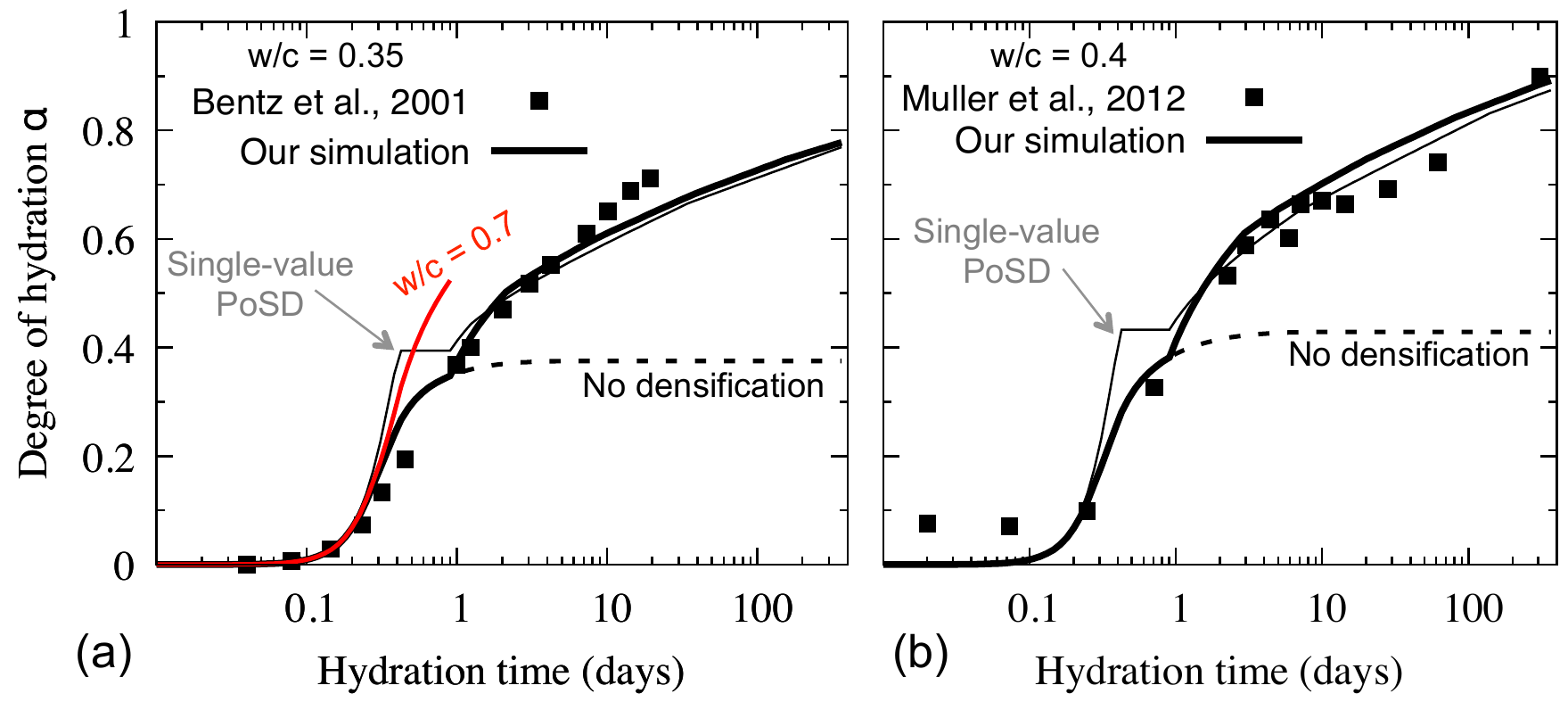}}
\caption{Calibration of model parameters to match experiments of degree of hydration evolving with time. (a) Low-alkali cement paste with $w/c = 0.35$~\cite{bentz2001influence} and (b) white cement paste with $w/c=0.4$~\cite{muller2014characterization}. The dashed lines show the effect of not allowing for C--S--H densification in the simulations ($\eta = \eta_{min}$ always). The thin lines show the effect of using a single-value initial size distribution of the capillary pores instead of a power-law distribution. Subfigure (a) shows also a prediction for a paste with $w/c = 0.7$.}
\label{figCalib}
\end{figure}

Once $\eta_{min}$, $\eta_{max}$, and $t_{peak}$ are fixed, the growth rate constant $G_{max}$ entirely controls the degree of hydration at the end of early hydration, when $t=t_{d0}$. The above-mentioned experiment by Bentz et al.~\cite{bentz2001influence} indicates $\alpha=0.4$ at one day of hydration. $G_{max}$ can thus be found iteratively by simulations, until the the desired $\alpha$ at 1 day is obtained. Such simulations however require an initial distribution of capillary pore sizes. This can be determined by running a first simulation assuming a single-value capillary PoSD, which provides $\overline D = 1.75$ $\mu$m for the paste in Bentz et al., and then, for the subsequent simulations, by using a power law distribution between $D_{min} = 1$ $\mu$m and $D_{max} = 10\overline D = 17.5$ $\mu$m. Choosing $D_{max}$ in the order of tenths of micrometres is supported by experimental measurements on pastes with $w/c$ around 0.4 \cite{diamond2000mercury}; pores of up to 100 $\mu$m have been recorded only in presence of entrained air, but they would be unsaturated and thus not contributing to hydration nor self-desiccation. In this way we find $G_{max} =  0.396$ $\mu$m hr$^{-1}$, which provides the early hydration kinetics $\alpha(t)$ in \figname\ref{figCalib}.a.

If all the other parameters are fixed, the densification rate constant $k$ controls the evolution of hydration over long time scales. In particular, in the experimental results by Muller et al.\cite{muller2014characterization}, who studied a paste with $w/c=0.4$ and Blaine fineness of the dry powder $S_s \approx 400$ m$^2$ kg$^{-1}$, the paste reached $\alpha \approx 0.9$ after one year of hydration in sealed conditions. Also for this, $\overline D = 2$ $\mu$m is determined assuming a single-value initial capillary PoSD, and then the simulations are carried out using an initial power law distribution of pores with diameters between $1$ and 20 $\mu$m. In this way one finds $k = 0.0083$ hr$^{-1}$, which leads to the long-term $\alpha(t)$ in \figname\ref{figCalib}.b.

\begin{table}[h!]
\centering
\caption{Calibrated parameters to be used in all simulations.}
\begin{tabular}{c c l} 
 \hline
Symbol 						&  Value       & Units   \\ [0.ex] 
 \hline  \\[-2ex]
 $\eta_{min}$ 			& 0.195			&  - \\ 
 $\eta_{max}$ 				& 0.74	 	&  -  \\
$t_{peak}$			&  10		& hr   \\
$G_{max}$		&  0.396		& $\mu$m hr$^{-1}$   \\
$t_{d0}$		&  1		& day   \\
$k$		&  0.0083		& hr$^{-1}$   \\
 \hline
\end{tabular}
\label{tabParam}
\end{table}

The parameters that all subsequent simulations will use for all the results in \secname\ref{secResults} are summarised in \tabname\ref{tabParam}. \figname\ref{figCalib} shows that:
\begin{itemize}
\item The kinetics resulting from the presented calibration is realistic, with hydration that continues over long time scales sustained by  the densification of the C--S--H.
\item The $w/c$ ratio has a limited effect on hydration kinetics during the first day, especially before the peak of hydration rate $\dot{\alpha}$ at $t\approx t_{peak} = 10$ hours in our simulations. This is a known feature of cement pastes \cite{masoero2014reaction}. 
However, due to the heuristic nature of the proposed hydration model, it is recommended not to rely on the present calibration to address $w/c$ ratios that are much different from the 0.35 -- 0.4 considered here;
\item Assuming a single-valued initial PoSD instead of a power-law PoSD has limited impact and the effect is only relevant to early hydration.
\end{itemize}

The experimental data in \figname\ref{figCalib}.a were only used to calibrate the model for the first day of hydration, whereas those in \figname\ref{figCalib}.b were used only to calibrate the late hydration ($t\rightarrow 365$ days). Therefore the late hydration in \figname\ref{figCalib}.a and the early hydration in \figname\ref{figCalib}.b are actual predictions validating the hydration model for the purpose of this paper.  {\color{black} Finally, it is worth noting that if one assumed non-cylindrical capillary pores (\eg spherical or slit pores in \figname\ref{figModel}), the values of the fitted parameters in \tabname\ref{tabParam} would probably be different, although most likely in the same order of magnitude because the surface-to-volume ratio of the pores is linked to fixed attributes of the paste (water to  cement ratio and specific surface area of the cement grains). Furthermore, once the $\alpha(t)$ relationship in \figname\ref{figCalib} is fitted, all the subsequent relationships between $\alpha$ (or time) and self-desiccation or sorption isotherms, which are the main focus of this manuscript, would be unchanged.}

\section{Results}\label{secResults}

This section shows the model predictions of five properties: (i) self desiccation, \viz decrease of internal relative humidity with hydration time, (ii) average C--S--H gel density, (iii) evolution of water volume in different pore categories, (iv) evolution of gel plus capillary pore size distribution, and (v) water sorption isotherms. The discussion of the results will follow on the effect of water-to-cement ratio, cement powder fineness, and curing temperature. The effect of considering \vs neglecting densification of the C--S--H gel will be discussed; the latter case assuming  $\eta_{min} = \eta_{max} = 0.655$, as in Ref.~\cite{masoero2014reaction}, \ie the gel forms immediately at intermediate density between the so-called ``low-density'' and ``high-density'' C--S--H in mature pastes, as typically assumed in available models of cement hydration. Simulation results will be compared to experiments on low-alkali cement pastes at room temperature and pressure, from three main sources:

\paragraph{\textbf{Muller et al.\cite{Muller201399,muller2014characterization}}} Sealed hydration of white portland cement with alkali content below 1\%w, Blaine fineness of the dry powder $S_s \approx 400$ m$^2$ kg$^{-1}$, and three water-cement ratios ($w/c=0.32$, 0.4, and 0.48). The hydrating pastes were monitored using $^1$H NMR, which provided the temporal evolution of different pore categories and an estimation of the evolving average density of the C--S--H gel. K\"{o}nigsberger et al. \cite{konigsberger2016densification} also analysed these NMR data obtaining a $w/c$--independent relationship between average C--S--H density and so-called ``specific precipitation space'' (volume of gel-plus-capillary pores filled with water divided by the sum of that same water-filled pore volume and the solid C--S--H volume). Muller et al.~also measured the self-desiccation of their paste with $w/c = 0.4$.

\paragraph{\textbf{Bentz et al.\cite{bentz2001influence}}} Sealed hydration of low-alkali ($<0.3$\%w) portland cement powder with $w/c=0.35$ and a range of Blaine finesses $S_s$ between 212 and 643 m$^2$ kg$^{-1}$. The temporal evolution of internal humidity iRH was measured alongside the evolution of chemical shrinkage. The latter was used to estimate the degree of hydration $\alpha(t)$ previously shown in \figname\ref{figCalib}.b.

 \paragraph{\textbf{Jensen and Hansen\cite{jensen1999influence}}} Sealed hydration of a white portland cement with alkali content of 0.55\%w Na$_2$O eq, $w/c = 0.3$, and Blaine fineness $S_s = 410$ m$^2$ kg$^{-1}$. The temporal evolution of iRH was monitored, and already shown in \figname\ref{figIntro}.a.

 {\color{black} The parameters of the initial pore size distributions resulting from \eqnames\ref{eqConstrV} and \ref{eqConstrS} for the combinations of $w/c$ and $S_s$ of the above-mentioned experimental samples are shown in \tabname\ref{tabPSDpar}. 
 
 \begin{table}[h!]
\centering
\caption{{\color{black} Parameters of initial pore size distributions for the main experimental samples considered in this manuscript. $N$, for the single-valued pore size distribution (PoSDI), is the number of capillary pores per unit mass of unhydrated cement grains.}}
\setlength{\tabcolsep}{9pt}
\begin{tabular}{c c c |c c| c c} 
 \hline
 			&		&		&\multicolumn{2}{c|}{Single PoSD}			&  \multicolumn{2}{c}{Power-law PoSD}  \\ 
Sample	   	& $S_s$ (m$^2$ kg$^{-1}$)		&$w/c$	&N ($10^{10}\;$g$^{-1}$) 	&$\overline{D}$ ($\mu$m)		&N ($10^{8}\;$g$^{-1}$)   	&$\zeta$ \\ [0.5ex] 
 \hline 
Muller et al. 	&400		 &$0.32$  			& 2.487		&  3.2 	&2.549    &4.117\\ 
Muller et al., 	&400		 &$0.4$  			& 1.592			&  4 	&3.071     &3.925\\ 
Muller et al. 	&400		 &$0.48$  			& 1.105			&  4.8 	&3.533    &3.802\\ 
Bentz et al. 	&212		 &$0.35$  			& 0.309			&  6.6 	&2.370     &3.639\\ 
Bentz et al. 	&643		 &$0.35$  			& 8.635			&  2.2 	&2.761     &4.650\\ 
Jensen \& Hansen 	&410		 &$0.3$  		& 3.047			&  2.9	&2.472     &4.218\\ 
 \hline
\end{tabular}
\label{tabPSDpar}
\end{table}

}

All simulations in this section, although referring to different pastes, use the same parameters calibrated in \secname\ref{secCalib}. This means that a quantitative agreement with the experiments may sometimes be beyond scope, but also that qualitative agreements will be true model predictions, not confounded by system-specific result-fits.

\subsection{Self-desiccation}
\begin{figure}[h]
\centerline{\includegraphics[width=1\textwidth] {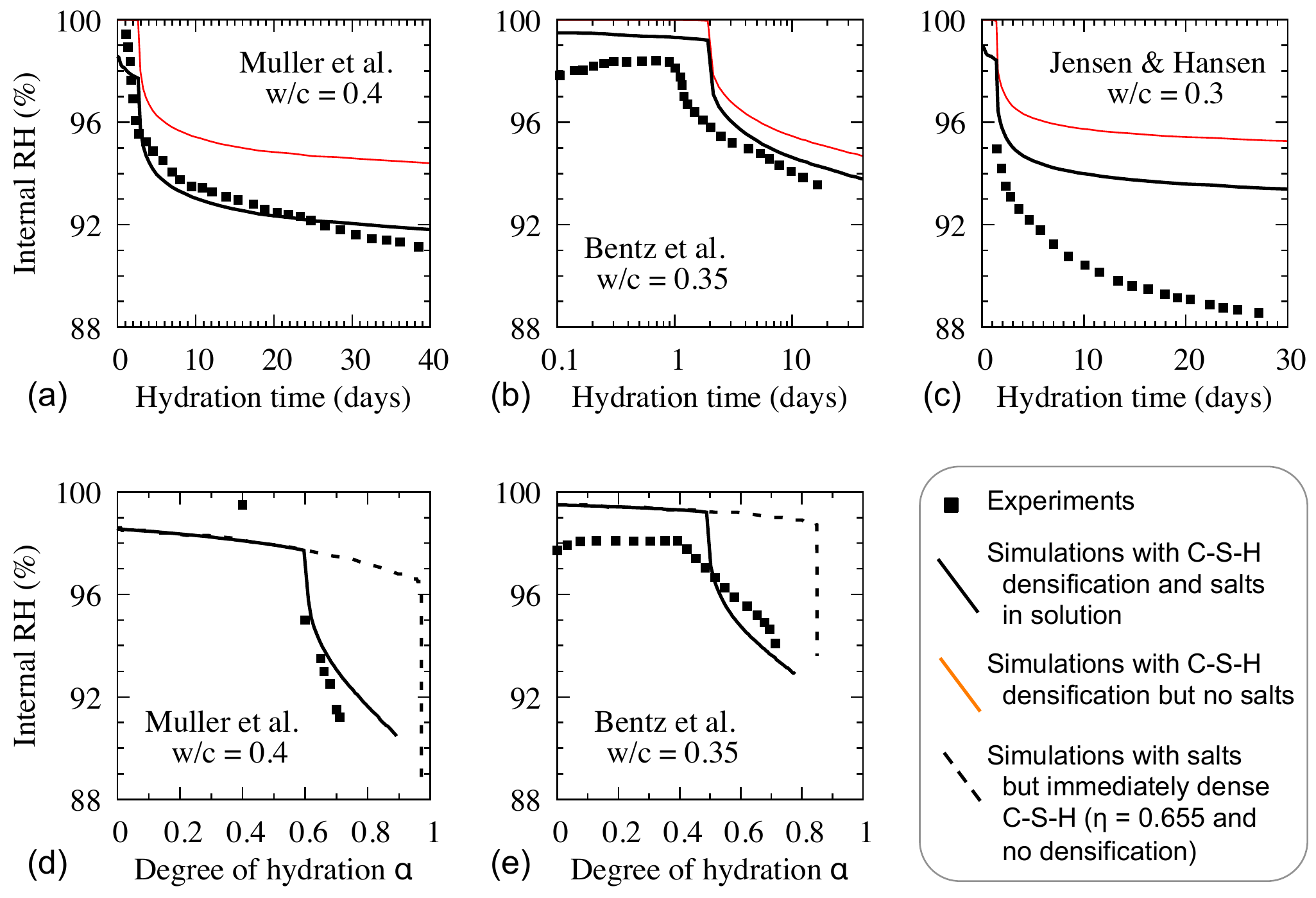}}
\caption{Self-desiccation predicted by our simulations for some of the experimental systems described at the beginning of this Results section.}
\label{figSelfD}
\end{figure}

\figname\ref{figSelfD} shows simulated self-desiccation of three pastes with slightly different $w/c$, obtained from cement powders with similar fineness ($S_s\approx 400$ m$^2$ kg$^{-1}$) and different, although always low, alkali content (0.3 -- 1\%w). \fignames\ref{figSelfD}.a-c focus on the temporal evolution of self-desiccation. The experiments show that an iRH of \ca 98\% is maintained during the first day of hydration, after which the iRH starts to decrease. %
The simulations predict an initial plateau of iRH, attributing it to the effect of ions in solution due to dissolution of salts, as agreed in the current literature \cite{lura2003autogenousTH}. The simulations predict a range of initial iRH at the plateau, between 98\% and 100\%, but an exact quantitative prediction of the iRH at the plateau is not straightforward using the simple approach based on Raoult's law. There may be other effects and processes determining the initial iRH, for example local concentrations of ions in solution. Such level of detail is beyond the scope of this manuscript, whose focus is rather on the onset of rapid self-desiccation after the initial plateau. 

The simulations in \fignames\ref{figSelfD}.a-c show that the occurrence of a sharp decrease of iRH after \ca one day of hydration is independent of dissolved salts, although considering the ions in solution gives a better quantitative agreement with the experiments. As hydration proceeds, the simulations overestimate the iRH or, that is the same, underestimate the self-desiccation, compared to the experiments. This may be adjusted by refining the model of C--S--H gel morphology evolution and using more detailed models for the location of water, \eg including surface adsorption in otherwise dry pores, entrained air during mixing, and ink-bottle effects. In this paper, however, a choice was made to keep such models as simple as possible, to show clearly that considering C--S--H gel densification is a necessary requirement in order to simulate even just qualitatively the early self-desiccation and evolution of water soprtion isotherms in the cement paste. To this end, \fignames\ref{figSelfD}.d and \ref{figSelfD}.e show the same iRH as in \fignames\ref{figSelfD}.a-c, this time plotted against the degree of hydration $\alpha(t)$ from \figname\ref{figCalib}, which referred to the same pastes considered here. Predictions are relevant to two simulations: both account for ions in solutions, but one considers that the C--S--H gel forms immediately as a dense and non-densifying phase with solid volume fraction  $\eta = 0.655$. The simulations with densifying C--S--H capture the drop of iRH at $\alpha \approx 0.4$, whereas those without densification are far off the experimental data. This shows that salts in solution alone are not sufficient to explain the experimentally observed self-desiccation.  Furthermore, in the simulations without gel densification, $G_{max}$ and $t_{peak}$ only set the time scale without any impact on the iRH($\alpha$) curves. Therefore the iRH($\alpha$) curve cannot be altered just by calibrating $G_{max}$ and $t_{peak}$ differently: the densification of an initially low-density C--S--H gel is necessary in order to obtain realistic iRH($\alpha$) relationships.

\subsection{C--S--H gel densification}\label{secDens}

\begin{figure}[h]
\centerline{\includegraphics[width=1\textwidth] {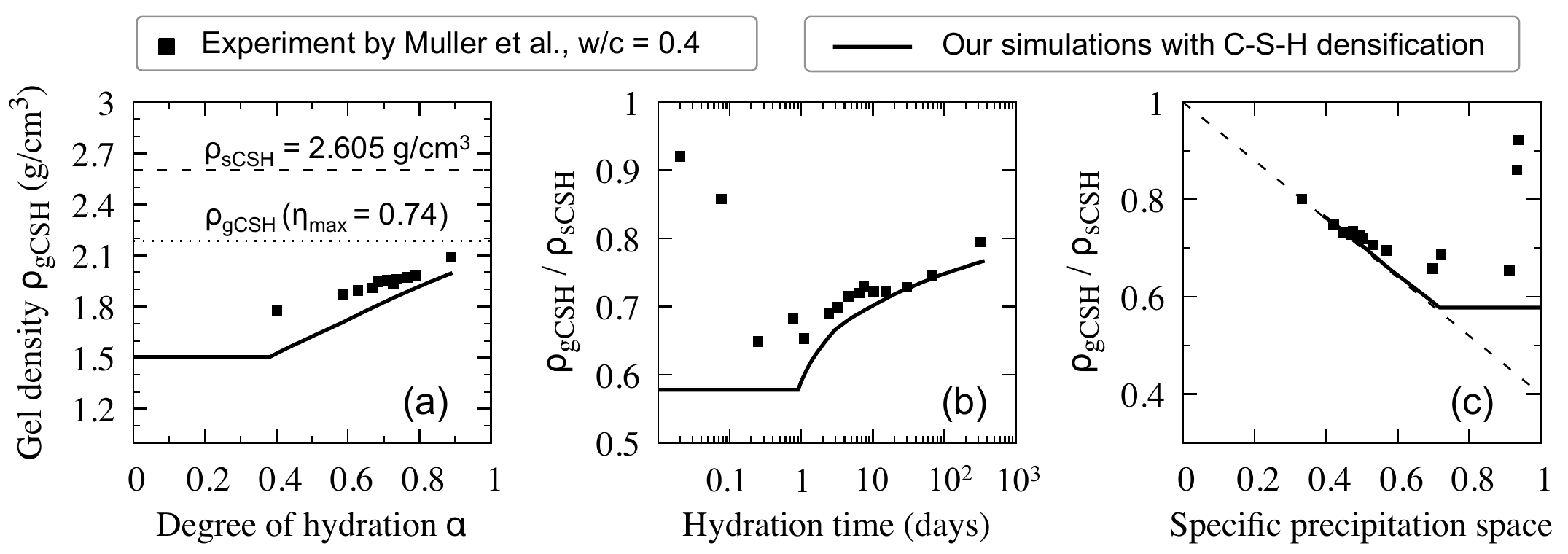}}
\caption{Evolution of C--S--H gel density $\rho_{gCSH}$ during hydration. All subfigures refer to the same experiment \cite{Muller201399}. (a) Evolution of  $\rho_{gCSH}$ with $\alpha$, compared to the the density of the solid C--S--H, $\rho_{sCSH}$, and to the maximum density of the C--S--H gel allowed in the simulations, when its solid volume fraction is $\eta_{max} = 0.74$. (b) Evolution of $\rho_{gCSH}$ with time and (c) with the specific precipation space, which decreases as hydration proceeds. The dashed line in (c) is a theoretical estimation from Ref.\cite{konigsberger2016densification} for a scenario in which the C--S--H gel has filled all the capillary space and $\rho_{gCSH}$ increases only due to densification.}
\label{figResDens}
\end{figure}

\figname\ref{figResDens} compares average C--S--H gel density $\rho_{gCSH}$ from the simulations and from $^1$H NMR experiments. The agreement is good at long time and large $\alpha$ (the specific precipitation space decreases with $\alpha$). The simulations instead underestimate $\rho_{gCSH}$ during early hydration ($\alpha \lesssim 0.4$, $t\lesssim 1$ day, specific precipitation space $\rightarrow$ 1). 

The origin of the early-age discrepancy in \figname\ref{figResDens} lies probably in the sub-micrometre morphology of the C--S--H gel. K\"{o}nigsberger et al.\cite{konigsberger2016densification} argue that the $\rho_{gCSH}/\rho_{sCSH}$ ratio during early hydration is large because the first C--S--H that forms is a non-porous solid, and only later  the C--S--H starts precipitating as a porous gel. Electron microscopy \cite{richardson2008calcium} shows that: (i) the first C--S--H displays so-called ``fibrillar'' or ``foil-like'' morphologies, probably related to a low Ca/Si ratio in solution \cite{tajuelo2015composition}; (ii) a porous gel phase appears later as hydration advances and the Ca/Si ratio in solution increases. The solid parts of fibrils and foils are non-porous but their strongly directional growth leaves behind a network of large nano-pores, with widths of hundredths of nanometres. 
$^1$H NMR might catalogue such large nano-pores as leftover capillary spaces rather than as gel pores, hence not considering them in $\rho_{gCSH}$. This scenario would reconcile the conflicting experimental observations of a large $\rho_{gCSH}$ during early hydration being accompanied nevertheless by early self-desiccation. The simulations do not capture the large initial $\rho_{gCSH}/\rho_{sCSH}$ ratio because they consider only porous-gel C--S--H morphologies, which are the only ones for which nanoscale simulation data of gel pore size distributions are available to date. Recent simulations are starting to reproduce fibrillar and sheet-like morphologies \cite{etzold2014growth} and to link morphology with solution chemistry \cite{shvab2017precipitation}. When computed, the nano-pore size distributions of such morphologies would enable addressing systematically the question of C--S--H gel density during early hydration.

\subsection{Evolution of different pore categories}

$^1$H NMR experiments by Muller et al.~\cite{Muller201399} quantify the amount of water in four pore categories and their evolution during hydration: (i) evapourable ``interlayer'' water adsorbed in the molecular structure of solid C--S--H, (ii) non-evapourable water chemically bound to hydration products such as CH and ettringite, (iii) ``gel'' water in the nano-pores of the C--S--H gel, more mobile than interlayer water but still significantly confined, and (iv) ``capillary'' water in pores that are sufficiently large to enable bulk-like mobility. In the simulations here, interlayer and bound water come directly from \eqname\ref{eqStoichio}, the gel water is all the water in the gel pores (both dense $\delta$ and loose $\lambda$ domains), and the capillary water is the water inside the part of cylinders in \figname\ref{figModel} not yet filled with hydration product. During early hydration (small $\alpha$) the simulations will overestimate the gel water and underestimate the capillary water compared to the experiments, because $^1$H NMR catalogues as capillary the water that the simulations from the presented model locate in large gel pores, \eg pores wider than $\sim$50 nm in the $\lambda$ gel domains. This is consistent with the underestimation of gel density previously discussed in relation to \figname\ref{figResDens}.

\begin{figure}[h]
\centerline{\includegraphics[width=0.5\textwidth] {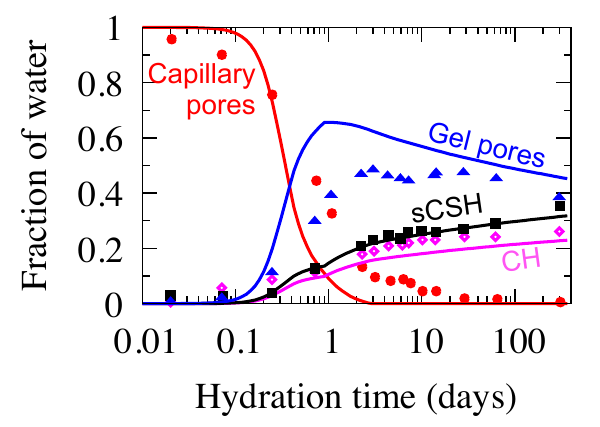}}
\caption{Evolution of water mass fraction in different pores: $^1$H NMR experiments by Muller et al.~\cite{Muller201399} on cement paste with $w/c=0.4$ (markers) and simulation results (lines).}
\label{figPoresSat}
\end{figure}

\figname\ref{figPoresSat} shows the evolution of water in different pore categories during hydration. The simulations match the experimental data on interlayer sCSH water, but this is simply because the model used this dataset to estimate the degree of hydration $\alpha(t)$ in \figname\ref{figCalib}.b. The other simulation results in \figname\ref{figPoresSat} however are direct predictions. For the bound water, only CH is considered as hydration product that might contain it, and despite this simplification the simulation results capture well the experimental data. For the gel water, the experiments show a rapid increase during the first day of hydration followed by a slow progressive decrease. The simulations capture this qualitative trend and provide insight into its origin: (i) the initial acceleration corresponds to the growth of low-density C--S--H with solid volume fraction $\eta_{min}$, which fills the capillary spaces; (ii) the subsequent decline starts when new gel stops forming, or forms in smaller amounts, because the capillary space is mostly filled by hydration product or desaturated. As expected and discussed above, the simulations overestimate indeed the volume of gel water during early hydration. After capillary space filling, hydration proceeds only by C--S--H densification and some gel pores start to get desaturated: both these processes reduce the amount of gel water. 

For the capillary water, the experiment in \figname\ref{figPoresSat} shows a rapid decrease during the first day of hydration, and then a slower decrease until \ca 100 days. Two processes control this trend: (i) reduction of capillary space due to the growth of hydration, and (ii) desaturation of capillary pores due to water consumption during hydration (at equilibrium, unreacted water preferentially fills the smallest pores, soon leaving the capillary pores dry). Therefore a zero value in \figname\ref{figPoresSat} does not mean that all capillary pores have disappeared: there can still be large but desaturated capillary pores. The simulations capture the fast decrease in signal during the first day of hydration, due to low-density C--S--H gel growing out rapidly.  As expect and discussed above, the simulations underestimate the amount of capillary water compared to the experiments. The simulations also overestimate the capillary water volume during very early hydration, at $t<0.01$ hours: this is probably because the experiment started with a paste that was already partially reacted, $\alpha(t=0)\approx0.1$, whereas the simulations assumed $\alpha(t=0)=0$ (see \figname\ref{figCalib}.b).

\subsection{Evolution of PoSD}

\begin{figure}[h]
\centerline{\includegraphics[width=0.5\textwidth] {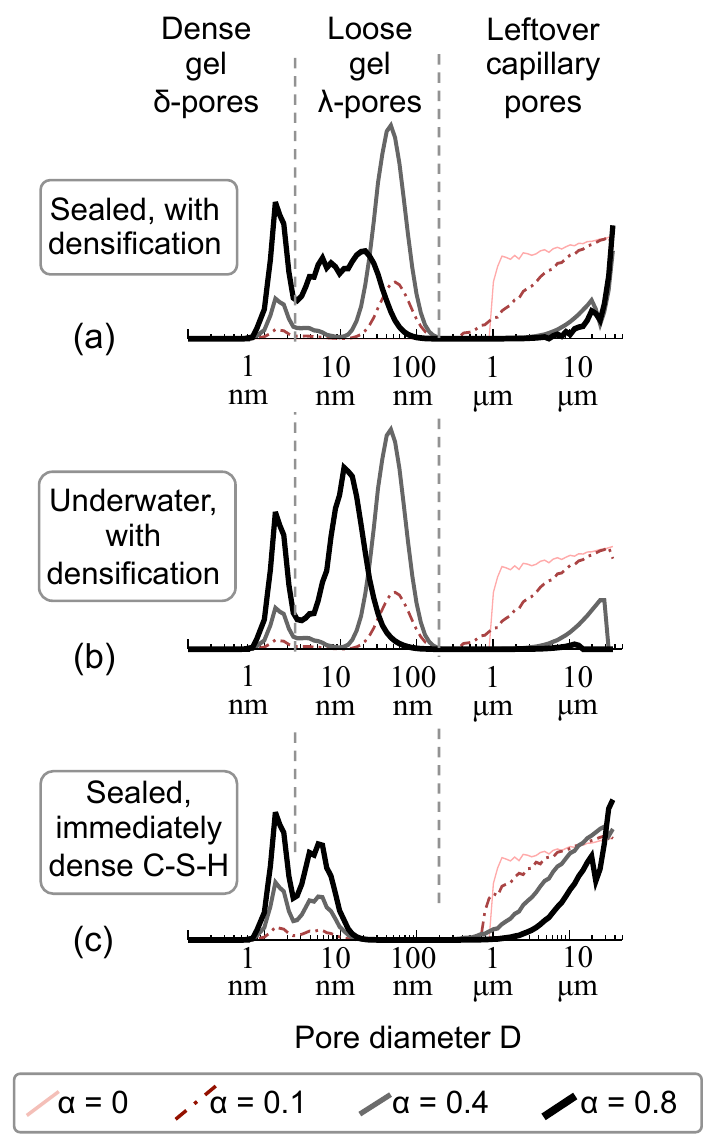}}
\caption{ { \color{black}Simulated pore size distributions of gel-plus-capillary pores, evolving with $\alpha$. The simulations refer to the Muller et al.~paste with $w/c=0.4$, analysed in the previous sections. (a) Results for a sample hydrating in sealed conditions, as in the original experiment, and for hydration products that densify as proposed in this paper. (b) Effect of changing hydration from sealed to underwater. (c) Strong difference in PoSD emerging if the densification of the hydration product is not considered. The vertical axis shows the quantity $\frac{1}{V_{p,tot}(\alpha)} \cdot\frac{dV_p(D)}{dlog_{10}D}$, where $V_p(D)$ is the volume of pores with diameter smaller than $D$, and $V_{p,tot}(\alpha)$ is the total volume of gel-plus-capillary pores at degree of hydration $\alpha$ (hence a unit area under each curve).} }
\label{figPoSD}
\end{figure}

The simulated size distributions of gel and capillary pores for the Muller et al.~paste with $w/c=0.4$ are shown in \figname\ref{figPoSD}.a as functions of the degree of hydration $\alpha$. When $\alpha = 0$, the PoSD is the power-law distribution from \secname\ref{secMethods}, for which 1 $\mu$m is taken as minimum initial pore width. This distribution aimed to mimic flocculation, thus small capillary pores are more abundant in number than large pores. Nevertheless the PoSD at $\alpha = 0$ in \figname\ref{figPoSD}.a increases with $D$ because the volume of pores scales as $D^3$ and because the $\mathrm{d\,log_{10}}D$ increment entails that the range of diameters corresponding to each ordinata in the plot increases with $D$.

When $\alpha = 0.1$, \viz during the first few hours of hydration, the overall volume and average size of the capillary pores decrease.
Meanwhile gel pores with $D$ in the 1-3 nm and 10-300 nm ranges appear. \figname\ref{figPoSD} indicates three ranges of pores attributing: the PoSD at $D<3$ nm to the dense domains $\delta$ of the C--S--H gel, the PoSD at $3<D<300$ nm to the loose $\lambda$ gel, and the PoSD at $D>300$ nm to leftover capillary pores, \viz the portions of the cylinders in \figname\ref{figCalib} that are still not filled by hydration product. Strictly speaking, the simulations allow for all pores $(\delta$ and $\lambda$ gel domains as well as capillary pores) to extend over the whole range of possible $D$. However, for the gel volume fractions considered here ($0.195 < \eta < 0.74$), the $\delta$-gel, $\lambda$-gel, and leftover capillary pores dominate indeed the PoSDs in the ranges shown in \figname\ref{figPoSD}.

When $\alpha = 0.4$, which corresponds to $t\approx 1$ day, \figname\ref{figPoSD}.a shows that the volume of capillary pores has drastically diminished, although some vary large capillary pores remain, persisting even at $\alpha = 0.8$. These are capillary pores that got desaturated and therefore cannot sustain further growth of C--S--H in them (the gel already in them, however, can still densify as long as it is saturated). 
This is consistent with the simulation result in \figname\ref{figPoresSat} showing no water in capillary pores when $t \ge 3$ days. These large leftover capillary pores are important for sorption and transport properties. The C--S--H gel at $\alpha = 0.4$ is largely dominated by pores in the loose $\lambda$ domains, because the model imposed $\eta = \eta_{min}$ and no densification during the first day of hydration. The volume of pores in the dense $\delta$ gel domains increased compared to the PoSD at $\alpha = 0.1$, but this is only due to the volume of new C--S--H gel that has grown in the capillary spaces, and not yet to densification.

When $\alpha = 0.8$, \ie nearly at complete hydration, the PoSD associated to gel pores in \figname\ref{figPoSD}.a has changed significantly. The relative abundance of pores in the dense $\delta$ domains increased at the expense of pores in the loose $\lambda$ domains: this is now mostly due to C--S--H densification, because the amount of new gel forming in capillary spaces at $\alpha>0.4$ is very small. The PoSD in the $\lambda$ domains changes qualitatively from the relatively sharp peak at $\alpha \le 0.4$ to a broad plateau between $D \approx 3$ nm to $D\approx 30$ nm. The plateau is determined by two competing processes: on one hand the saturated C--S--H gel densifies, shifting the average size of pores of the $\lambda$ domains (the $\lambda$-peak in the PoSD) towards a smaller $D$; on the other hand, the lastly formed C--S--H gel, which has the smallest $\eta$, starts to get desaturated, cannot densify anymore, and thus contributes with a $\lambda$-gel PoSD that cannot evolve towards a smaller average $D$.

\figname\ref{figPoSD}.b refers to the same system as in \figname\ref{figPoSD}.a, but this time the C--S--H can always densify and can grow in any capillary pore, irrespective of the saturation state. This mimics hydration underwater, if one assumes that water can always access all pores. In this case the capillary pores disappear completely as $\alpha \rightarrow 1$. Furthermore, one of the two competing effects causing the $\lambda$-gel plateau in \figname\ref{figPoSD}.a does not take place (namely the arrest of densification due to C--S--
H gel desaturation). As a result the densification of all the C--S--H gel leads to a single well-defined peak for the $\lambda$-pores, which moves towards smaller average $D$ values as $\alpha$ increases. Underwater conditions do not change significantly the PoSD during early hydration, $\alpha \le 0.4$, because most of the capillary and all gel pores are saturated also in sealed conditions.

\figname\ref{figPoSD}.c considers again the sealed conditions as in \figname\ref{figPoSD}.a, but this time the C--S--H gel is assumed to form immediately as a rather dense phase with solid volume fraction $\eta = 0.655$ and is not allowed to densify (like in the dashed curves in \figname\ref{figSelfD}). The resulting PoSDs are strongly affected by this assumption. The capillary pore volume reamins much larger compared to \fignames\ref{figPoSD}.a and \ref{figPoSD}.b. This is already evident at $\alpha = 0.1$ and persists at $\alpha = 0.8$, with the very last peak of PoSD at $D\approx40$ $\mu$m due to desaturated capilllary pores, and the nearby peak at $D\approx20$ $\mu$m due to the initial capillary space that is still largely unfilled by the hydration product. The corresponding evolution of gel porosity is trivial: because densification is not allowed, the pore structure within the gel cannot change, the PoSD remains always identical in shape, and only its height increases as new C--S--H gel grows into the capillary pores. The double peak in the gel PoSD in \figname\ref{figPoSD}.c is because at $\eta = 0.655$ the description of the C--S--H gel in \figname\ref{figDens} implies a coexistence of dense and loose domains, the latter with a rather small average $D$ but still larger than that of the dense domains.

\subsection{Water sorption isotherms}

\begin{figure}[h]
\centerline{\includegraphics[width=1\textwidth] {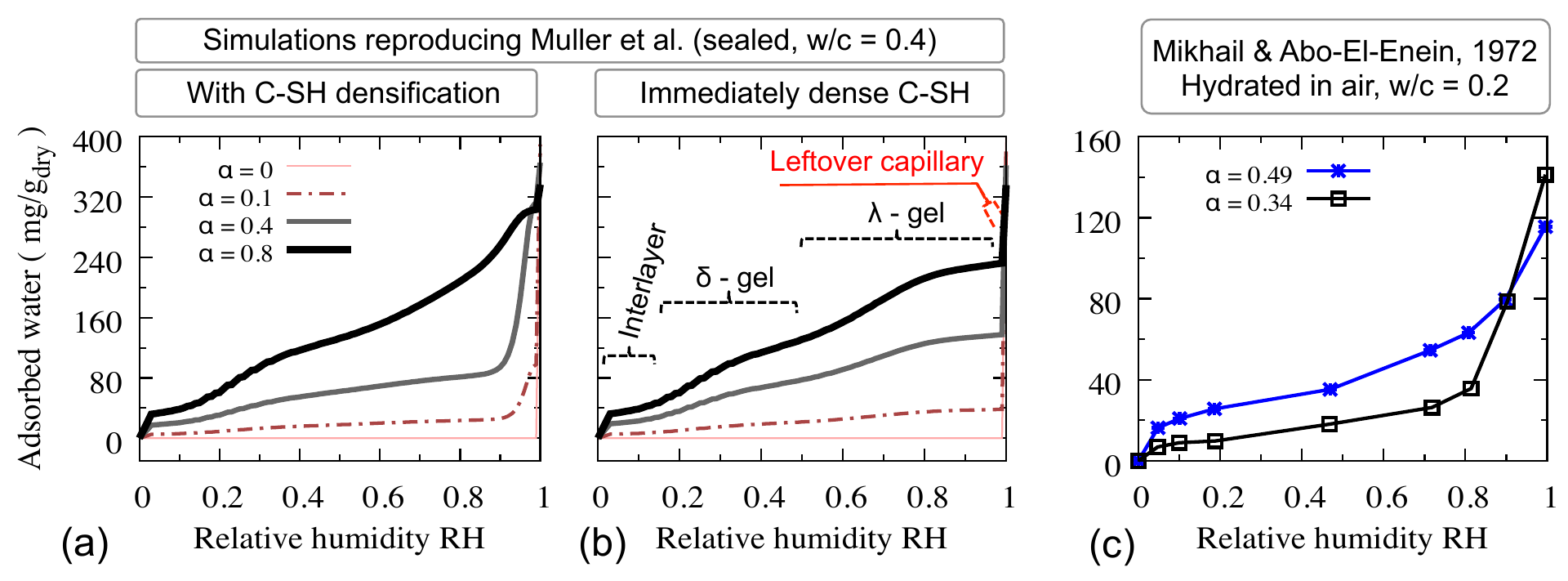}}
\caption{(a) Simulated sorption isotherms as a function of $\alpha$ for the Muller et al.~paste with $w/c=0.4$. (b) Same simulations as (a), but assuming that the C--S--H forms immediately with $\eta=0.655$ and does not densify. (c) Experimental isotherms for a Type I portland cement paste with $w/c = 0.2$ and Blaine fineness $S_s = 620$ m$^2$ kg$^{-1}$\cite{Mikhail1972}. (b) also highlights which pore categories dominate adsorption in different RH ranges.}
\label{figResIso}
\end{figure}

The pore size distributions in \figname\ref{figPoSD} lead to the water sorption isotherms in \figsname\ref{figResIso}.a and \ref{figResIso}.b. This section discusses only qualitative features of the simulated isotherms, evidencing the impact of C--S--H densification. Quantitative results will be discussed later, in relation to the effect of the water-to-cement ratio.

\figname\ref{figResIso}.a shows simulated isotherms when C--S--H densification is allowed. When $\alpha \rightarrow 0$, a limit that cannot be studied experimentally, adsorption in the simulations can only occur in capillary spaces with $D\ge 1$ $\mu$m, for which the Kelvin equation predicts saturation at RH$\approx 1$ (see \figname\ref{figIntro}.d).
 As $\alpha$ increases from 0 to 0.4, the volume of water adsorbed in leftover capillary pores at $RH\approx 1$ decreases, because the capillary pores are partially filled by hydration product. Correspondingly, an increasing volume of water is adsorbed in gel pores, which account for most of the isotherms between RH $\approx0.2$ and 0.99, as shown in \figname\ref{figResIso}.b. The gel PoSD in \figname\ref{figPoSD}.a shows indeed an increasing volume of both small $\delta$-gel and large $\lambda$-gel pores, but their amount relative to the other, and thus their relative impact on adsorption, do not evolve with $\alpha$ yet because C--S--H densification has not begun yet and all the outer C--S--H gel has same morphology and solid volume fraction $\eta_{min}$. 
As $\alpha$ increases further from 0.4 to 0.8, the isotherms in \figname\ref{figResIso}.a highlight instead a twofold effect of C--S--H gel densification: (i) more water gets adsorbed in $\delta$-gel pores (RH$\lesssim 0.5$) as $\alpha$ increases at the expense of water in $\lambda$-pores; (ii) adsorption in $\lambda$-pores occurs sharply at $RH\approx0.9$ when $\alpha = 0.4$ and becomes more gradual over a range of RH between 0.5 and 1 as $\alpha\rightarrow0.8$, reflecting the plateau in $\lambda$-gel pores PoSD in \figname\ref{figPoSD}.a.

Overall, \figname\ref{figResIso}.a shows that the isotherms are almost entirely determined by the nano-pores in the C--S--H gel, and therefore by the gel morphology. Large capillary pores, leftover from hydration, contribute only to the total amount of water that can be adsorbed at full saturation, achievable experimentally only by forcing liquid water into a sample. A closer look at \figname\ref{figResIso}.a shows that the total water at saturation decreases with $\alpha$, as expected because the total porosity decreases and the dry mass increases with $\alpha$.

\figname\ref{figResIso}.b refers to the same simulations as in \figname\ref{figResIso}.a, but this time assuming that the C--S--H forms immediately with solid volume fraction $\eta = 0.655$ and does not densify further (the corresponding PoSD was in \figname\ref{figPoSD}.c). For RH $\lesssim 0.5$, the isotherms in \figname\ref{figResIso}.b are very similar to those in \figname\ref{figResIso}.a, because adsorption in this humidity range is dominated by $\delta$-gel domains whose morphologies are identical in the two simulations, irrespective of C--S--H densification. On the other hand, for $RH\gtrsim 0.5$ the isotherms in \figname\ref{figResIso}.b indicate less adsortpion compared to those in \figname\ref{figResIso}.a, because there is much less $\lambda$ gel in the former (see \figname\ref{figPoSD}.c). Indeed the isotherms in \figname\ref{figResIso}.b are almost flat at RH between 0.8 and 0.99, whereas experimental isotherms usually show a significant slope (see \figname\ref{figResIso}.c). The simulations with C--S--H densification in \figname\ref{figResIso}.a instead agree qualitatively with the experiment, such as those in \figname\ref{figResIso}.c, predicting a finite slope at large RH due to C--S--H gel domains that are not fully dense and that contain pores with widths of tenths of nanometres. Furthermore, in absence of gel densification, the relative amount of adsorption in $\delta$ and $\lambda$ gel pores stays constant during hydration and therefore, unlike the experiments, the simulations without densification in \figname\ref{figResIso}.b do not predict any evolution of the shape of the isotherm with $\alpha$. All this indicates that a good description of gel morphology is crucial in order to predict realistic water sorption isotherms and their evolution during hydration.

\subsection{Effect of water-to-cement ratio}

\begin{figure}[h]
\centerline{\includegraphics[width=.8\textwidth] {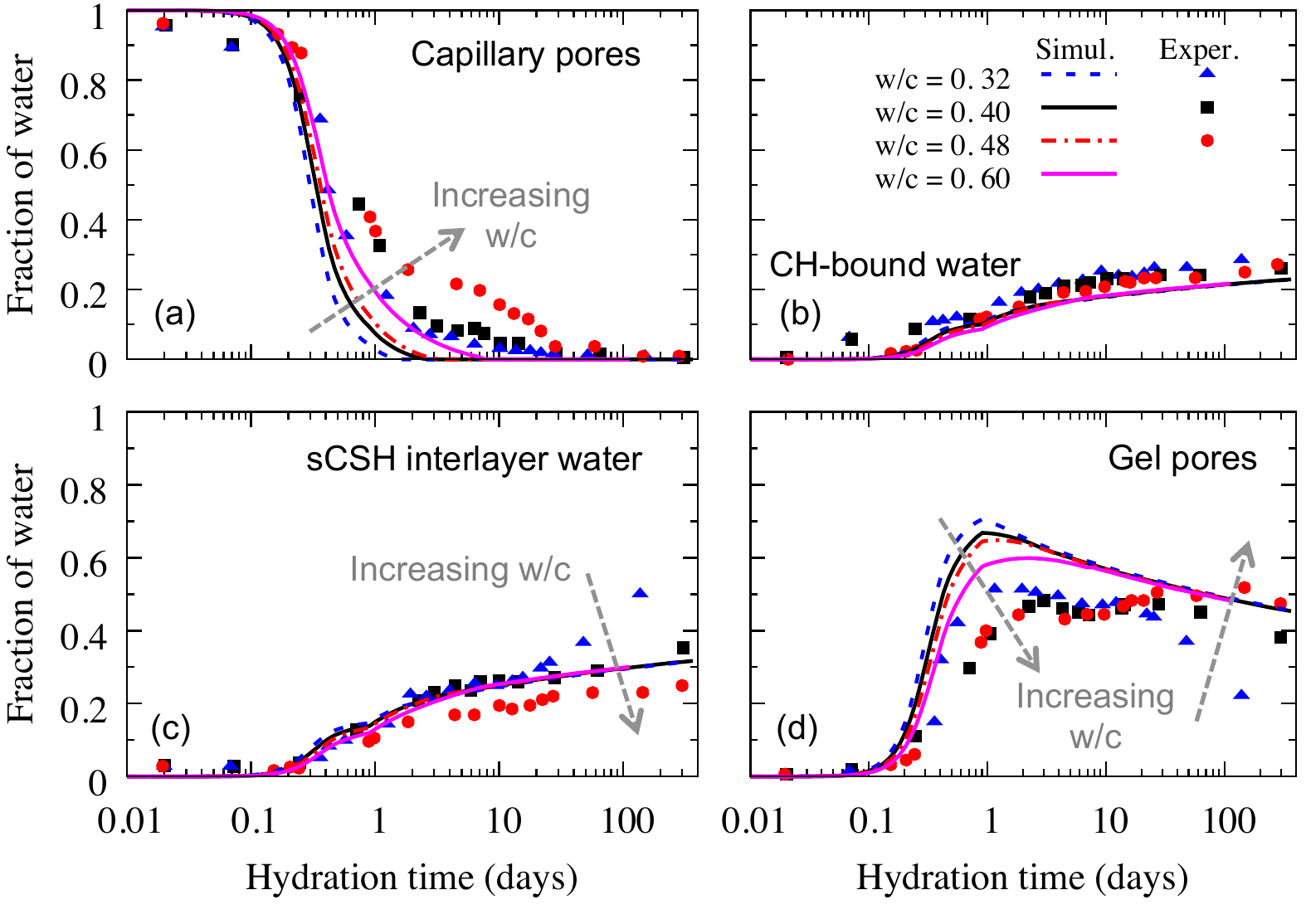}}
\caption{Effect of $w/c$ on the presence of water in different pore categories during hydration. $^1$H NMR experiments by Muller et al.\cite{muller2014characterization} and our simulations.}
\label{figPoresWC}
\end{figure}

\figname\ref{figPoresWC} shows the effect of $w/c$ on the amount of water in different pore categories. The simulations in \figname\ref{figPoresWC}.a predict correctly that the capillary pores in pastes with higher $w/c$ take longer to get filled with hydration product and/or desaturate. The experiments show a similar trend, but with a larger volume of capillary water than in the simulations. As previously mentioned, this may be due to NMR grouping together large gel pores ($D\gtrsim 50$ nm) and leftover capillary space. Consistently, the simulations in \figname\ref{figPoresWC}.d predict more gel pore water than the experiments. They also correctly predict that the fraction of water in gel pores during the first 1-3 days of hydration is smaller for pastes with higher $w/c$: this is because high-$w/c$ pastes have more of saturated capillary pores, as shown in \figname\ref{figPoresWC}.a.

\begin{figure}[h]
\centerline{\includegraphics[width=.5\textwidth] {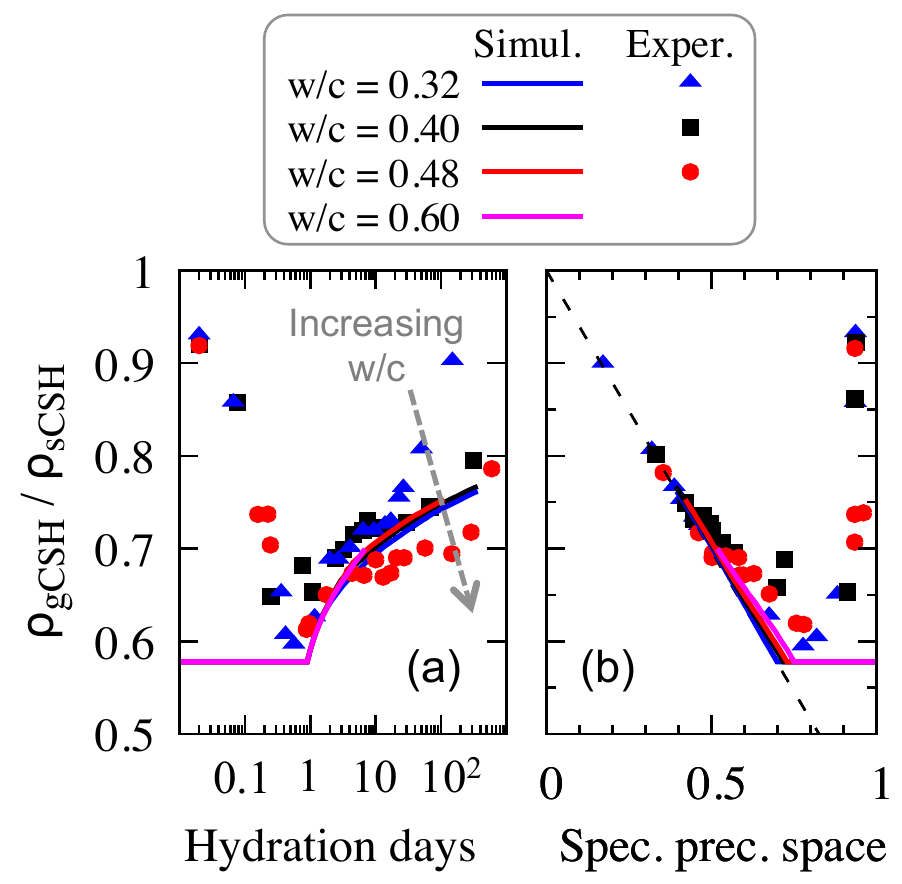}}
\caption{Effect of $w/c$ on the presence of water in different pore categories during hydration. $^1$H NMR experiments by Muller et al.~\cite{muller2014characterization} \vs predictions from our simulations. The lack of agreement during early hydration (first day and large specific precipitation space) was already discussed in relation to \figname\ref{figResDens}.}
\label{figResDensWC}
\end{figure}

The simulations in \figname\ref{figPoresWC}.b capture well the evolution of chemically bound water, which is independent of $w/c$. The evolution of interlayer pores in the solid C--S--H in  \figname\ref{figPoresWC}.c is more problematic, as the simulations do not predict any effect of $w/c$ whereas the experiments display two features: (i) starting from \ca 3 days of hydration, the paste with highest $w/c = 0.48$ has less fraction of interlayer water compared to pastes with lower $w/c$; (ii) after \ca 20 days, the interlayer water fraction increases markedly in the paste with smallest $w/c=0.32$. These features are mirrored by the experimental results in \figname\ref{figPoresWC}.d, showing that the fraction of gel pore water increases with the $w/c$ during late hydration (\eg 100 days) . This all suggests that the C--S--H gel in pastes with a high $w/c$ is less dense, \ie has a smaller average volume fraction $\eta$ (see \figname\ref{figResDensWC}). 

To capture the relationship between $\eta$ and $w/c$ in the experiments in \figname\ref{figPoresWC}, K\"{o}nigsberger et al. \cite{konigsberger2016densification} proposed to relate $\eta$ to the so-called specific precipitation space (SPS), defined at beginning of this section. Their result is shown in  \figname\ref{figResDensWC}.b, with the experimentally measured C--S--H gel densities $\rho_{gCSH}$ now independent of the $w/c$ ratio. The model proposed in the current work does not impose an explicit relationship between $\eta$ and $w/c$ and as a result, unlike the experiments, it does not not predict the impact of $w/c$ on the temporal evolution of $\eta$ in \figname\ref{figResDensWC}.a. To fit the experiments, one would need to introduce explicit dependence on $w/c$ in the $k$ factor of the densification rate in \eqname\ref{eqDenst2}. The model would also have to assume a stoichiometry in \eqname\ref{eqStoichio} that changes during hydration, in order to preserve the $w/c$-independent evolution of chemically bound water in \figname\ref{figPoresWC}.c while allowing for more (or less) solid C--S--H to precipitate and densify the gel. A possible rationale for $w/c$-dependent densification rates could be to relate them to the rate of ion diffusion: shorter diffusion paths in low-$w/c$ pastes may lead to faster densification. However, part of the discrepancy with the experiments might be due to the NMR experiments classifying as interlayer some of the water that the simulations locate instead in very small gel pores, \eg $D\lesssim 1-2$ nm. More experimental data are needed in order to clarify this point that, although interesting, is not central to the discussion of the impact of C--S--H densification and $w/c$ on self-desiccation and water sorption isotherms.

\begin{figure}[h]
\centerline{\includegraphics[width=.5\textwidth] {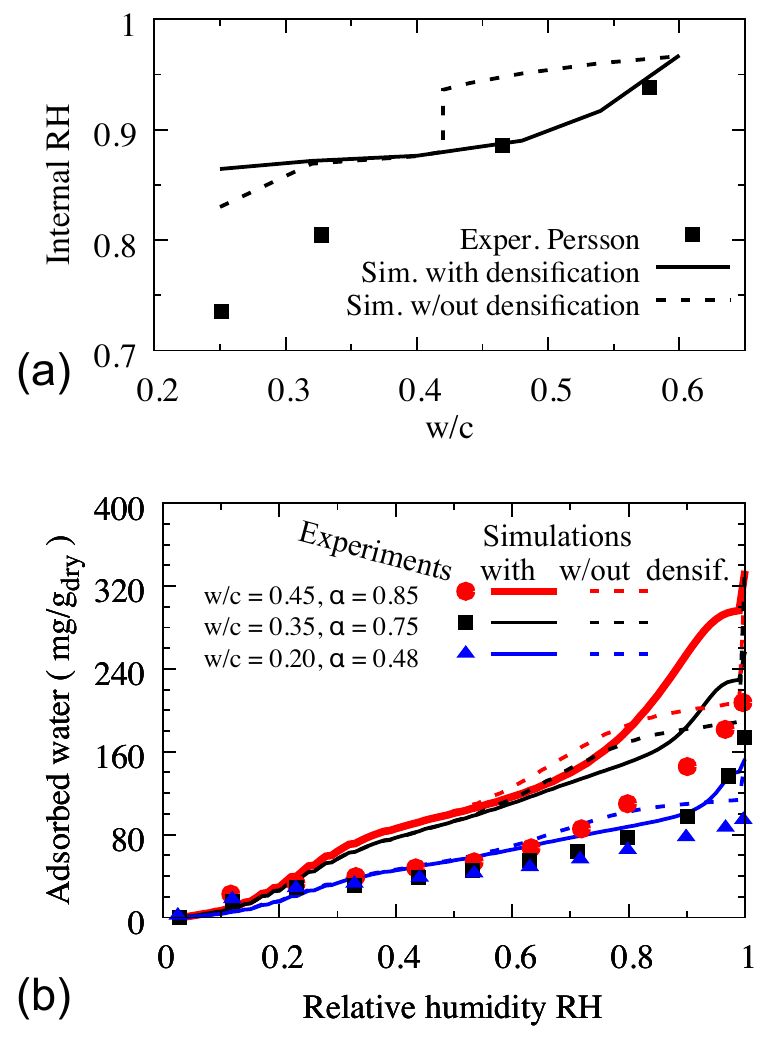}}
\caption{(a) Self-desiccation of concrete after 15 months of hydration: experiments from Persson \cite{persson1998pozzolanic} and our simulations. The simulations without densification assume that the C--S--H gel is a non-densifying phase with $\eta = 0.655$. (b) Water sorption isotherms for cement pastes with different $w/c$: experiments from Baroghel-Bouny \cite{baroghel2007water} after 2 years of hydration and simulations with same $w/c$ and degrees of hydration $\alpha$ as in the experiments.}
\label{figSelfDIsoWC}
\end{figure}

\figname\ref{figSelfDIsoWC}.a shows experimental results on self-desiccation of concrete after 15 months of hydration \cite{persson1998pozzolanic}. The concrete contains portland cement with \ca 2\%w of alkali sources, $w/c$ between 0.25 and 0.6, and Blaine fineness $S_s = 325$ m$^2$ kg$^{-1}$. The simulations in \figname\ref{figSelfDIsoWC}.a refer to pastes with same $w/c$, $S_s$, and alkali content as in the experiments, but with hydration kinetics calibrated as in \secname\ref{secCalib}, thus not fitted to the specific experiments. For $w/c \ge 0.48$, the simulation with C--S--H densification capture well the relationship between self-desiccation and $w/c$, whereas not considering densification leads to a significant overestimation of the internal RH, which does not decrease much  going from $w/c \approx 0.6$ to 0.45 (only a small decrease  due to greater concentrations of ions in solution, because the dense C--S--H gel is unable to fill the capillary space and to start getting desaturated, not even at full hydration, $\alpha = 1$).

At low $w/c$, all the simulations in \figname\ref{figSelfDIsoWC}.a predict a higher iRH compared to the experiments. This might be due to multiple factors, \eg additional alkali in the aggregates of the concrete or difficulty to maintain sealing over 15 months. There are however other interesting features in the results at low $w/c$. Unlike the experiments, the simulations without densification predict a sharp drop of iRH at $w/c \approx 0.42$, which is the theoretical limit below which full hydration becomes impossible. This means that, at some point during hydration, the dense gel fills all the saturated capillary space in pastes with low $w/c$, thus the iRH becomes controlled by the gel pores. In simulations without densification the size of the gel pores is independent of $w/c$, therefore the iRH for $w/c\lesssim 0.42$ should be constant, except for some decrease due to higher concentration of salts in solution at low $w/c$. The simulations with C--S--H densification show a similar decrease of iRH with $w/c$ below 0.42, but predict less salts-induced self-desiccation at very low $w/c = 0.25$. The reason is that the average gel density at 15 months in the simulations with densification is smaller than in the simulations without densification, because the densifying gel can get desaturated during hydration and thus stop densifying further. The simulations with densification therefore attain a lower degree of hydration, more water remains in the pores, and thus the salt concentration remains lower than in the simulations without densification.

\figname\ref{figSelfDIsoWC}.b shows experimental water adsorption isotherms \cite{baroghel2007water} for CEM I - 52.5 cement pastes with different $w/c$ ratios, alkali content below 1\%w, and Blaine fineness $S_s\approx 330$ m$^2$ kg$^{-1}$. The pastes were cured for 2 years in sealed conditions, reaching different degrees of hydration as shown in the figure. The paste with $w/c = 0.2$ also contained 10\%w of silica flour. The reference dry condition, following Baroghel-Bouny, is at 3\% RH. According to \figname\ref{figIsoInter} this corresponds to the solid C--S--H in the simulations being still 30\% saturated: this is accounted for when normalising the mass of adsorbed water by the mass of the dry sample in \figname\ref{figSelfDIsoWC}.b. All the simulations capture the qualitative trend of increasing adsorption with increasing $w/c$, which simply reflects the larger total porosity at high $w/c$. C--S--H densification, as already shown in \figname\ref{figResIso}, has a significant impact on the shape of the isotherms at RH $\gtrsim 0.6$. The experiments in \figname\ref{figSelfDIsoWC}.b display an increasing gradient at RH $\gtrsim 0.6$, predicted correctly by the simulations with C--S--H densification whereas simulations without densification predict a decreasing gradient.

In quantitative terms, the simulations in \figname\ref{figSelfDIsoWC}.b over-predict the experimental saturation. This may be improved by refining the description of the C--S--H gel morphology and by considering the actual chemistry of the cement paste in more details (this latter affecting for example the dry mass on the vertical axis). Entrained air during mixing and ink-bottle effects causing hysteresis may also play a role \cite{pinson2015hysteresis}. On the other hand, the overestimation may be more intrinsic to the adopted model of capillary condensation. Water sorption experiments do not achieve full saturation at RH = 1, \viz in a fog environment; full saturation requires immersion in water with hydraulic head. Simulations assuming Kelvin equation and cylindical pores predict instead full saturation at RH = 1. The Kelvin equation is based on average curvatures of water menisci, and the complex pore network morphology of realistic pastes can lead to null average curvatures already in small spaces (consider for example a capillary bridge between two spherical grains, with positive curvature on the plane of the grains and negative curvature on the plane perpendicular to the line connecting the centres of the grains). Simulating capillary condensation in complex nanopore networks is an active field of research \cite{kierlik2002adsorption}, and direct nanoscale simulations of water adsorption may be needed in order to improve the quantitative agreement with the experiments.

\subsection{Effect of cement powder fineness}

\begin{figure}[h]
\centerline{\includegraphics[width=.5\textwidth] {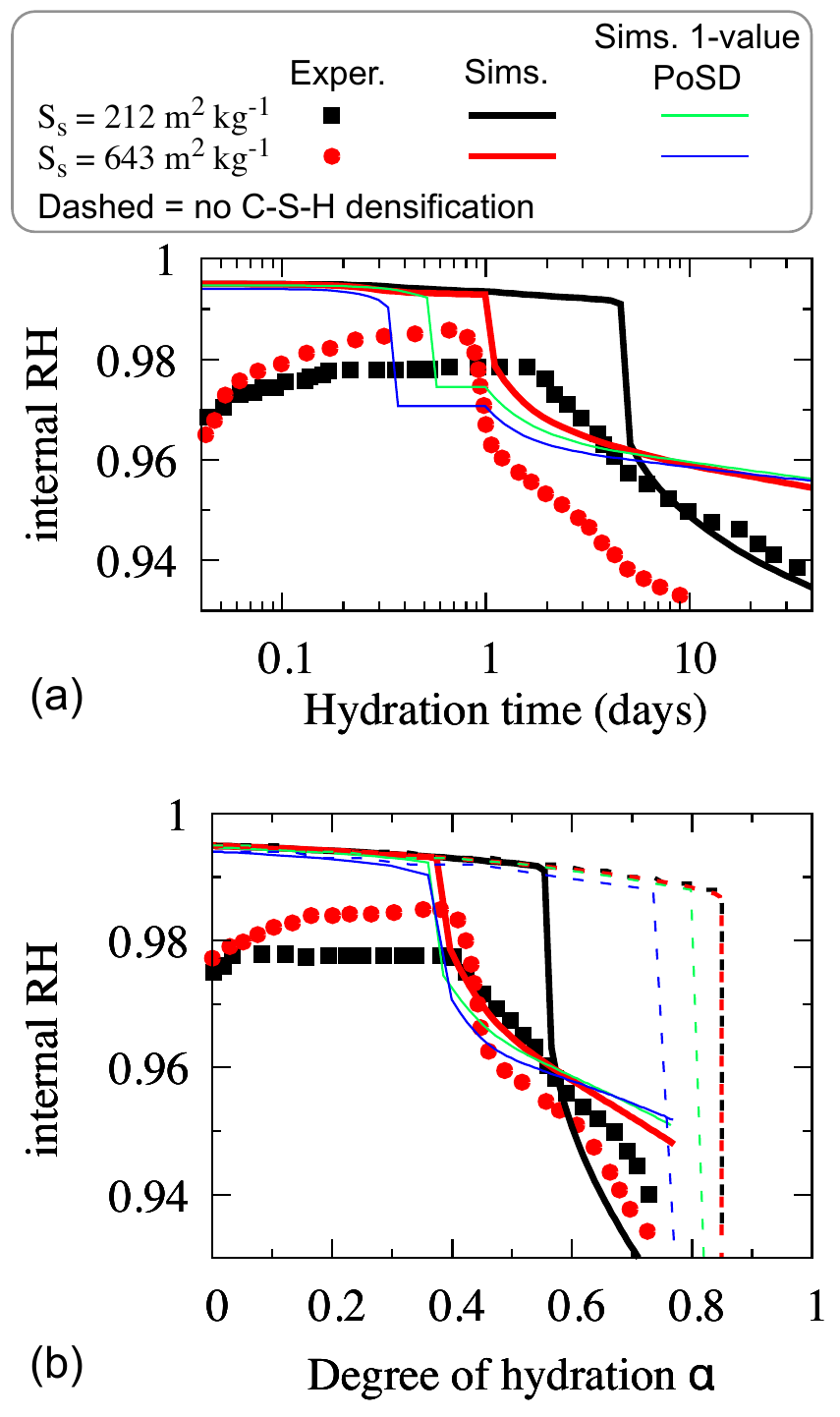}}
\caption{Self-desiccation of pastes from Bentz el al.\cite{bentz2001influence}, with different specific surface area of the cement powders (same chemistry as the pastes in \figname\ref{figSelfD}, but different Blaine fineness here): evolution with (a) time and (b) degree of hydration. Also shown are the effects of assuming a non-densifying C--S--H gel with solid fraction $\eta = 0.655$ and of assuming a single-valued initial distribution of capillary pore sizes.}
\label{figFine}
\end{figure}

The experiments in \figname\ref{figFine}.a show that pastes obtained from fine powders start to desiccate earlier than pastes from coarser powders. The simulations presented in this paper predict a similar trend. However, in \figname\ref{figFine}.b the simulations predict also that pastes from fine powders should start to self-desiccate at smaller $\alpha$ compared to pastes from coarse powder. The experiments instead indicate that the iRH($\alpha$) relationship does not depend on the fineness of the cement powder. Furthermore, the simulations predict a crossover at \ca 5 days and $\alpha \approx 0.55$, with the iRH of pastes from coarser powders becoming smaller than the iRH of pastes from finer powders. The experiments do not display such crossover. The difference between simulations and experiments depends in part on the choice of initial capillary pore size distribution. \figname\ref{figFine} shows result from simulations assuming a single-valued initial capillary PoSD, \viz all cylinders in \figname\ref{figModel}.b have same initial size $\overline{D}$. The results in this case still capture the earlier onset of desiccation of pastes from fine powders in \figname\ref{figFine}.a (the quantitative fit of the time scale is lost because the hydration model was calibrated assuming a power law initial PoSD in \secname\ref{secCalib}) and predict also an iRH($\alpha$) relationship that does not depened on the fineness of the powder. There is scope for future investigations of how to best model the capillary PoSD, but hereafter the discussion focuses on the impact of C--S--H densification on self-desiccaiton.

\figname\ref{figFine}.b shows that if the C--S---H is not allowed to densify, the onset of self desiccation is predicted at unrealistically large $\alpha \approx 0.82$, irrespective of the fineness of the powder. Using a single-valued initial PoSD instead of a power-law one reduces slightly the $\alpha$ at the onset of self-desiccation, but this $\alpha$ becomes dependent on the powder fineness (unlike the experiments) and does not decrease below 0.7 (whereas experimental self-desiccation occurs already at $\alpha \approx 0.4$).

\subsection{Effect of curing temperature}

Experimental results show that cement pastes, cured to the same degree of hydration, develop less strength if the curing is done at high temperature instead of room temperature. Small angle neutron scattering shows that high-temperature curing leads to coarser C--S--H gel structures \cite{jennings2007multi}, or in other words, a shift towards looser $\lambda$ rather than denser $\delta$ gel domains. 

Temperature is not an explicit parameter of the presented model, but is implicit in the growth and densification rates and in the relationship between density $\eta$ and gel pore size distribution. It is assumed here that: (i) temperature affects only the growth and densification rates, and (ii) these two rates are controlled by different limiting processes with distinct activation energies. For the growth rate $G$, differential calorimetry experiments during early hydration indicate an activation energy of \ca 50 kJ mol$^{-1}$ \cite{Thomas2012}. C--S--H densification, instead, is important during late hydration, when the kinetics is largely controlled by diffusive processes with lower activation energy of \ca 20 kJ mol$^{-1}$ \cite{longsworth1954temperature,bavzant2015interaction,rahimi2017cement}. Processes with higher activation energy are more affected by temperature changes, thus it is reasonable to consider that an increase of curing temperature impacts more the growth rate than the densification rate. Therefore one can model higher curing temperatures by increasing the ratio between $G_{max}$ and $k$ in \eqnames\ref{eqG1} and \ref{eqDenst2}.

\begin{figure}[h]
\centerline{\includegraphics[width=.5\textwidth] {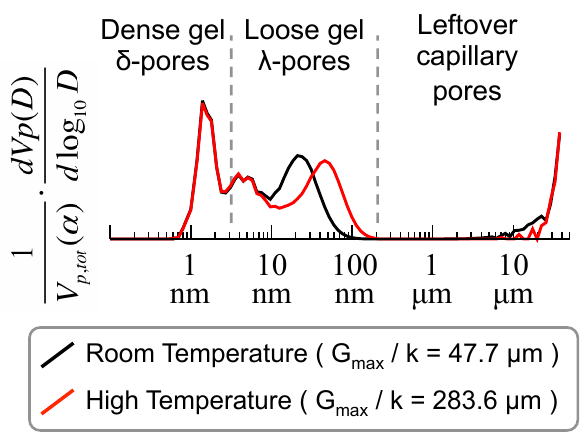}}
\caption{Simulated pore size distributions for same paste hydrated at two different temperatures, modelled \via the $G_{max}/k$ ratio (see text). The simulations refer to the paste studied in Muller et al.~with $w/c = 0.4$, hydrated to a degree $\alpha = 0.85$.}
\label{figPoSDtemp}
\end{figure}

\figname\ref{figPoSDtemp} shows that a higher curing temperature (larger $G_{max}/k$ ratio) leads to a larger average pore size in the loose $\lambda$ domains of the gel. The PoSD in the dense $\delta$ domains is negligibly affected, whereas the remaining volume of capillary pores is slightly smaller at higher curing temperature. Multi-scale poromechanics suggests that the rigidity percolation threshold, \viz the minimum solid fraction below which the porous paste is not mechanically rigid anymore, starts from 0 at the macroscale and increases as the length-scale of the observation is reduced \cite{bernard2003multiscale}. This implies that adding pore volume as nano-pores is more detrimental for the mechanical performance than adding the same volume as macro-pores. Therefore a change of porosity like the one in \figname\ref{figPoSDtemp}is likely to result in poorer mechanical strength for pastes cured at higher temperatures.

\section{Conclusion}
The work presented in this paper shows that a detailed description of the nano-scale morphology of the C--S--H gel and of its evolution during hydration is essential in order to predict correctly the onset of self-desiccation and the evolving shape of water sorption isotherms during early hydration. This was achieved using a simple model of cement hydration and focussing on the evolution of the multiscale pore size distribution of the paste. The key novelty of the model is to leverage state-of-the-art nanoscale simulations of C--S--H gel formation, to obtain a constitutive description of the evolving morphology of the C--S--H gel and in particular its gel pore size distribution. 

The simulations captured early self-desiccation and sorption isotherms, but some experimental results are still to be fully reconciled. In particular: the density of the C--S--H gel during the first days of hydration, the quantitative agreement with water sorption isotherms, and the effect of water-to-cement ratio on the density of the C--S--H gel. This is not surprising, given the simplicity of our hydration model. {\color{black}Maintaining this simplicity has been an intentional choice aimed at proving that the ability of a model to capture early self-desiccation and sorption isotherms lies in information on the nanostructural evolution of the C--S--H gel, and not on chemical or microstructural complexity at larger length scales. Advanced simulators of cement hydration and microstructure development already exist, \eg CEMHYD3D, HydratiCA, $\mu$ic, and Hymostruc \cite{Bullard2011}, thus it is recommended that future works follow the example of this paper on how to implement nanostructure-related information into advanced simulators, rather than improving the hydration-specific component of the simple model employed here. This would also allow studying pastes with more complex chemistry, for example in high-performance concrete containing silica, which can undergo harsher self-desiccation and associated damage compared to the low alkali cement pastes considered in this paper \cite{lura2003autogenous}.
}

Important challenges such as the long-term disposal of nuclear waste and the sustainability and resilience of the infrastructure network, require a new understanding of how nanoscale degradation mechanisms impact the macroscale properties of cementitious materials. Such an understanding is growing, supported by: (i) new experimental characterizations of the nanoscale morphology of the C--S--H gel, and its dependence on chemical composition and curing conditions \cite{tajuelo2015composition}, (ii) growing capabilities of nanoscale simulations to incorporate chemical kinetics and to model the experimental results \cite{shvab2017precipitation}, and (iii) an increasing awareness that macroscale engineering models need inputs from micro and even nano scale studies \cite{bavzant2015interaction,pinson2015hysteresis,do2013numerical}. Having shown some of the potential benefits of combining nanoscale simulations with macroscale models, this work strengthens the synergy between nanoscale cement science and macroscale engineering mechanics.

\bigskip
\textbf{Acknowledgments}
E.M. acknowledges the support of the TU1404 COST Action, EU Framework Programme Horizon 2020. E.M. also thanks the Matua Campus of Politecnico di Milano, for supporting his stay at Politecnico di Milano in April 2017. The work of G.C. was supported under NRC grant NRC-HQ-60-14-G-0003.

\section*{References}

\bibliographystyle{unsrt}
\bibliography{bibliocement}


\end{document}